\documentstyle[12pt,aaspp4]{article}
\newcommand{\ul}{\underline{\hspace{40pt}}}

\begin{document}

\title{Radiative Hydromagnetic Shocks in Relativistic Outflow Sources}


\author{Jonathan Granot$^1$ and Arieh K\"onigl$^2$}

\affil{$^1$Racah Institute, Hebrew University, Jerusalem 91904,
Israel; jgranot@nikki.fiz.huji.ac.il \newline $^2$Department of Astronomy \& Astrophysics and
Enrico Fermi Institute, University of Chicago, 5640 S. Ellis
Ave., Chicago, IL 60637; arieh@jets.uchicago.edu}
\setcounter{footnote}{2}
\begin{abstract}
  We calculate the structure of a relativistic shock wave in which the
  internal energy of the shocked fluid is radiated away on a time
  scale much shorter than the characteristic shock propagation time.
  The shock is assumed to move through a uniformly magnetized, neutral
  plasma consisting of protons, electrons, and positrons, and
  allowance is made for the possible production of electron/positron
  pairs in the shock itself.  The radiation mechanism is taken to be
  synchrotron and inverse-Compton emission (the latter involving
  both synchrotron-produced and externally supplied seed photons)
  by the electrons and positrons.  We simplify the discussion by
  considering a shock in which the magnetic field is transverse
  to the direction of propagation and focus attention on the
  properties of the radiative zone that forms behind the shock
  transition. In particular, we investigate the possibility that
  the compression induced by the cooling of the gas amplifies
  the magnetic pressure until it reaches (and ultimately
  exceeds) equipartition with the thermal pressure (which, in
  turn, limits the overall compression). We show that, if a
  significant fraction of the postshock thermal energy is deposited in
  the electron/positron component, then a considerable portion of the
  emitted radiation will come from regions of strong magnetic field
  even if the field immediately behind the shock transition is well below
  equipartition. This field amplification mechanism may be
  relevant to the production of synchrotron flares in blazars,
  miniquasars, and gamma-ray burst sources. We consider the
  latter application in some detail and show that this process
  may play a role in the prompt $\gamma$-ray and possibly also
  the optical ``flash'' and radio ``flare'' emission, but
  probably not in the afterglow.
\end{abstract}


\section{Introduction}

Propagating relativistic shock waves are the leading candidates for
the origin of the emission from gamma-ray bursts (GRBs) and their
afterglows (e.g., Piran 1999, 2000; M\'esz\'aros 2000) as well as of the
spectral flares and apparent superluminal motions exhibited by the
blazar class of active galactic nuclei (AGNs; e.g., Ulrich, Maraschi,
\& Urry 1997) and their ``miniquasar'' Galactic counterparts (e.g.,
Mirabel \& Rodr\'{\i}guez 1999; \cite{KSS}). The shocks are thought to
be associated with relativistic, jet-like
outflows that are driven from the vicinity of a compact object: a
massive black hole in blazars, a stellar-mass black hole in
miniquasars, and a stellar-mass black hole or a rapidly rotating
neutron star in GRBs.
\footnote{In at least some GRBs there are, however,
indications that the outflow may have a
fairly large opening angle or even be spherical.}
In all cases except GRB afterglows, the shocks
are believed to be {\em internal}, possibly resulting from the
collision of ``shells'' (the products of nonsteady ejection episodes)
inside the jet (e.g., Rees 1978; \cite{FMN}; \cite{SP97b}). In the
case of GRB afterglows, the most common interpretation is in terms of
shocks that form at the interface between the relativistically
outflowing material and the surrounding medium (with the
bulk of the observed emission arising in the ``forward'' shock that
propagates into the ambient medium,
whereas the ``reverse'' shock that is driven into the ejecta
gives rise to an early ``optical flash'' and associated ``radio flare'').
The radiation in all three types of sources is inferred to
be nonthermal, with the dominant emission mechanisms most commonly
invoked being synchrotron and inverse Compton. In particular, both the
$\gamma$-ray burst emission and the radio--through--X-ray afterglow
radiation in GRBs, as well as the radio--through--X-ray emission in
blazars and the radio--through--optical radiation in microquasars, has
been attributed to synchrotron radiation (or, in certain models of
GRBs, a closely related mechanism; e.g., \cite{SU}, \cite{MV00},
\cite{LB01}). In the case of blazars, the measurement of high
($\gtrsim 10\%$) optical linear polarization in some objects directly
supports the synchrotron interpretation (e.g., Angel \& Stockman
1980). In the case of GRBs there are already two confirmed detections
of optical polarization (Covino et al. 1999; Wijers et al. 1999; Rol
et al. 2000), which, despite the relatively low measured values ($\sim
1-3\%$), appear to be consistent with a synchrotron origin.  In both
blazars and microquasars the case for synchrotron radiation is
supported also by measurements of radio polarization.

Synchrotron radiation requires the presence of charged particles
(typically electrons or electron-positron pairs) with relativistic
random velocities that move in a magnetic field. The magnitude of the
energy densities of the random relativistic electron motions and of
the magnetic field are parametrized in terms of their ratios to the
total thermal energy density in the emission region by $\epsilon_e$
and $\epsilon_B$, respectively. In most cases, the values of these two
parameters cannot be accurately determined from the observational
data. However, in the case of blazars there are indications that the
emitting gas is in approximate equipartition between the particle and
magnetic energy densities (i.e., $\epsilon_e \approx \epsilon_B$ ---
e.g., Readhead 1994; Bower \& Backer 1998). This conclusion is also
consistent with models of the nonthermal flares in microquasars (e.g.,
Kaiser et al. 2000). In the case of GRBs, if the observed $\gamma$-ray
emission indeed represents synchrotron radiation, then considerations
of radiative efficiency (based on estimates of source energetics)
imply that $\epsilon_e$, and likely also $\epsilon_B$, are not much
smaller than 1 (e.g., \cite{SP97a}; Piran 1999).  In the case of GRB
afterglows, there have been a few instances in which data were
available for deriving more direct estimates of these parameters. In
particular, theoretical fitting of the broadband spectrum of the
afterglow of GRB 970508 has yielded $\epsilon_e \sim 0.1-0.6$ and
$\epsilon_B\sim 0.01-0.1$ (\cite{W&G99}; \cite{GPS99}; \cite{C&L00}).

The inference that $\epsilon_e$ and $\epsilon_B$ are not much smaller
than 1 in GRBs and at least some afterglows has important implications
to the nature of the emitting regions in these sources. A
comparatively large value of $\epsilon_e$ implies a high efficiency of
shock acceleration of relativistic electrons in an electron-proton
plasma or else a high electron-positron ($e^\pm$) pair content in the
postshock region (reflecting the composition of the preshock gas or
arising from copious pair production in the shock). A comparatively
large value of $\epsilon_B$ implies either a highly magnetized
preshock medium or else an efficient mechanism of field amplification
in (or behind) the shock. If $\epsilon_e \sim 1$ and the radiative
cooling time is much shorter than the dynamical time\footnote{The
dynamical time is the time for a significant expansion of the
emitting region, and is thus also the characteristic time for
adiabatic energy losses. It is typically of the order of the shock
distance from the origin divided by the instantaneous shock speed.}
(which, for synchrotron radiation, would require also $\epsilon_B$ not
to be too small), then the shock will be {\em radiative}: it will
contain a radiative cooling layer, where most of the internal energy
is radiated away, that will directly influence its dynamics.
Nonrelativistic, thermally emitting shocks of this type are prevalent
in the interstellar medium (e.g., \cite{D&M93}), but their properties
in relativistic, nonthermal sources have not yet received much
attention in the literature. 

A radiative shock wave generally consists of a ``shock transition,''
where the bulk of the kinetic energy of the preshock gas is
dissipated, and a ``radiative zone,'' where most of the dissipated
energy is radiated away. The energy dissipation scale is typically
much smaller than all other relevant length scales, so the shock
transition can be treated as an infinitesimal front at the upstream
end of the shock wave. The radiative zone, on the other hand, has a
finite width, corresponding to the cooling length over which the
postshock gas radiates away a significant fraction of the dissipated
energy. After passing through the shock transition, the gas is usually
compressed by a factor of a few (for Newtonian or mildly relativistic
shocks), and in the absence of significant radiation this is the
density enhancement that also characterizes the far-downstream (or
postshock) region. However, it is a well-established result in the
theory of interstellar shocks that the asymptotic compression can be
far larger if the shock is radiative. This is a consequence of the
fact that the total (thermal plus magnetic) pressure behind the shock
transition remains roughly constant, so that, as long as magnetic
effects are not important, the particle density $n$ increases
inversely with decreasing temperature $T$.

A related important result
for radiative interstellar shocks is that, if the gas has a
``frozen in'' magnetic field $B$ with a significant component ($B_{\perp}$)
perpendicular to the shock normal, then magnetic
stresses become progressively more important as $T$ decreases
behind the shock, a reflection of the fact that the magnetic
pressure, associated with $B_{\perp}$, scales as $n^2$. If the
preshock magnetic field is sufficiently strong, then the
magnetic pressure in the cooling layer will
reach equipartition with the thermal pressure at some point
within the radiative zone. Beyond this point the gas
will undergo little additional compression, and the cooling will
proceed at nearly constant density rather than at constant
thermal pressure (e.g., Hollenbach \& McKee 1979). Since the cooling time is
proportional to $1/n$ (a consequence of the fact that the
cooling function $\Lambda$, the energy radiated per unit volume
and per unit time, scales as $n^2$ for a thermally
emitting gas), the inhibition of further compression by the
magnetic stress results in a lower cooling efficiency in this part
of the radiative zone.

One can anticipate that the basic properties of radiative
hydromagnetic shocks in the interstellar medium (ISM) would carry over
to the relativistic, nonthermal regime. In particular, one can expect
a significant enhancement of the compression as well as of the
downstream magnetic field in a radiative shock.  The two cases are,
however, expected to differ in detail. In particular, the cooling
function of a synchrotron-emitting shocked gas satisfies $\Lambda
\propto B^2 n_e \propto n_e^3$ (assuming
a predominantly transverse magnetic field behind the shock), where
$n_e$ is the number density of electrons (and positrons). The explicit
density dependence of $\Lambda$ is thus stronger than in a thermally
emitting gas, which suggests that the dynamical influence of the
magnetic field on the structure of the shock would be even more
pronounced in the nonthermal case. This could have important
implications to astrophysical sources in which such shocks occur. For
one thing, it may provide a mechanism for amplifying the magnetic
field up to equipartition strengths in cases where it is initially
comparatively weak. Furthermore, as the processes of field
amplification and synchrotron emission are mutually reinforcing in the
radiative zone, this mechanism may naturally lead to synchrotron
emission regions that are characterized by a high value of
$\epsilon_B$. There is even the possibility, if inverse-Compton
cooling also plays a role, that the bulk of the emitted
synchrotron radiation would come from regions where the magnetic field
is already at (or near) equipartition.\footnote{Relativistic
radiative shocks have previously been considered in the context
of blast-wave models (\cite{BM}; \cite{CPS98}; \cite{CP99}). In
particular, Cohen et al. (1998) and Cohen \& Piran (1999)
obtained self-similar solutions for spherical blast waves that
have adiabatic interiors and that are bounded by fully (or partially) radiative
ultrarelativistic (or Newtonian) shocks. These studies
incorporated the efect of radiative losses in a parametric way
into the overall shock jump conditions. They did not,
however, include specific cooling mechanisms and did not
consider the detailed structure of the shock wave.}

In this paper we examine the aforementioned effects in a
quantitative manner by studying the properties of relativistic,
hydromagnetic, nonthermally emitting shocks. 
In \S \ref{shocktran} we describe the framework of our analysis and
calculate the shock jump conditions. The structure of the
radiative zone is considered in \S \ref{radzone}. In \S \ref{B1}
we discuss the circumstances under which radiative shocks of this
type are likely to occur in astrophysical sources, focusing on
the case of GRBs and their afterglows. Our conclusions are
presented in \S \ref{conclusions}.

\section{The Shock Transition}
\label{shocktran}

\subsection{Formulation of the Problem}
\label{formulation}

We analyze the steady-state propagation of relativistic, planar,
radiative shocks in a medium that contains a large-scale, ordered
magnetic field. As we noted in \S 1, radiative shocks can be divided
into a very thin ``shock transition'' and a ``radiative zone'' of
finite thickness. We assume, for simplicity, that all the energy
dissipation occurs in the shock transition and all the cooling in the
radiative zone. The assumption of a stationary flow throughout the
shock is justified by the fact that, by the definition of radiative
shocks, the radiative cooling time $t_{\rm rad}$ of the gas that
passes the shock transition is much shorter than the shock dynamical
time $t_{\rm dyn}$. The assumption of a plane-parallel flow is easily
justified for the shock transition, and we now argue that it is also
appropriate for the radiative zone.

The global shape of the shock is generally not planar. For example, if
the shock is triggered by a nearly isotropic energy release at the
source, as in supernova remnants and certain GRB scenarios, then it
may assume an approximately spherical shape. The instantaneous
distance of the shock front from the central source, $R$, is then
given by the radius. If the geometry is not exactly spherical, then
one needs to distinguish between the distance $R$ and the local radius
of curvature $R_c$. Typically, $R_c\sim R$ and both are much larger
than the width of the radiative zone. For example, in a relativistic
blast wave of Lorentz factor $\gamma \gg 1$, most of the energy and
swept-up matter are concentrated within a thin shell of width
$\Delta\sim R/\gamma^2$ in the observer frame, or
$\Delta'=\gamma\Delta\sim R/\gamma$ in the shock frame (e.g.,
\cite{BM}). The width of
the radiative zone is $\sim \Delta (t_{\rm rad}/t_{\rm dyn}) \ll
\Delta \ll R$ in the observer frame, or $\sim\Delta'(t'_{\rm
rad}/t'_{\rm dyn})=R'_c(t_{\rm rad}/t_{\rm dyn})\ll R'_c$ in the
shock frame (where $R'_c=R/\gamma$ is the local radius of curvature in
the shock frame).  In any reasonable scenario, $R'_c\gtrsim \Delta'$,
and even for  $R'_c\sim\Delta'$, the width of the radiative
zone would still be smaller than $R'_c$ by a factor $\sim t_{\rm
rad}/t_{\rm dyn}\ll 1$.
Given also that $R_c$ is not expected to
change considerably on time scales shorter than the dynamical time, it
is thus well justified to assume that the shock is planar when
calculating the properties of the radiative zone.

We distinguish between the shock rest frame, in which we measure the
time $t$, the distance behind the shock front $x$, and the fluid speed
$v=\beta c$, and the fluid rest frame (the frame comoving with the
local mean speed of the flow), in which we measure the thermodynamic
quantities (the particle number density $n$, the rest-mass density
$\rho$, the internal energy density $e$, the thermal pressure $p$,
and the enthalpy density $w=\rho c^{2}+e+p$) as well as the magnetic
field amplitude $B$. We simplify the treatment by assuming that
the field is {\em purely transverse}, $B = B_{\perp}$, i.e., it only has a
component that is perpendicular to the shock velocity and
parallel to the shock front. This is justified by the fact that,
upon traversing the shock, $B_{\perp}$ is amplified
by a factor equal to the downstream-to-upstream density ratio (which can be
quite large in a relativistic shock) whereas the parallel component remains
unchanged, so that the postshock magnetic field is typically almost completely
transverse.

\subsection{Shock Jump Conditions}
\label{SJC}

We denote quantities in the unshocked upstream
region ($x < 0$) by the subscript 1, whereas quantities immediately
behind the shock transition are denoted by the subscript
2. Quantities without numerical subscripts refer to the
radiative zone. The shock is
assumed to propagate in an electrically neutral plasma consisting of
protons, electrons, and possibly also positrons. We define the
parameter
\begin{equation}\label{chi}
\chi\equiv{n_{e^+}\over n_{e^-}}={n_{e^+}\over n_p+n_{e^+}}\ ,
\end{equation}
where $n_p$, $n_{e^-}$ and $n_{e^+}$ are the number densities of
protons, electrons, and positrons, respectively. We assume that the
number of protons is unchanged by the passage through the shock
transition, but we allow
the number of $e^\pm$ pairs to possibly increase in the shock
transition (although not during the subsequent transit through
the radiative zone).
Thus we have
\begin{equation}\label{n}
u_1 n_{p1}=u_2 n_{p2}=u n_p\ ,\quad\quad
\phi u_1 n_{e1} = u_2 n_{e2}=u n_{e}\ ,
\end{equation}
where $n_e=n_{e^-}+n_{e^+}$ is the combined number density of
electrons and positrons, $u=\gamma\beta$ is the proper speed (in
units of the speed of light $c$), and the 
parameter $\phi$ accounts for the possible production of pairs in the shock.
The constraint $\phi\geq 1$ implies that
\begin{equation}\label{chi_2}
\chi_2={\phi(1+\chi_1)-(1-\chi_1)\over\phi(1+\chi_1)+(1-\chi_1)}\geq\chi_1\ .
\end{equation}
From equations (\ref{chi}) and (\ref{n}) we infer
\begin{equation}\label{n_p2}
n_{p2}=\left({1-\chi_1\over 1+\chi_1}\right){u_1\over u_2}\, n_{e1}\ .
\end{equation}
With this parameterization it is straightforward to take the following
limits: (i) no pair production at the shock corresponds to $\phi=1$;
(ii) no protons corresponds to $\chi_1=\chi_2=1$; (iii) no pairs
upstream corresponds to $\chi_1=0$.

The pressure and energy density of
the magnetic field are given by $p_B=e_B=B^2/8\pi$, and hence the
total pressure and enthalpy density are $p_{\rm tot}=p+B^2/8\pi$ and
$w_{\rm tot}=w+B^2/4\pi$, respectively. In this paper we
focus on the case in which the fluid is ``cold'' upstream and
``hot'' downstream. The former assumption means that the 
thermal pressure and internal energy density of the upstream fluid can be
neglected ($e_1, \, p_1\ll \rho_1 c^2$), so that 
\begin{equation}\label{w1}
w_1 \approx \rho_1 c^2=n_{e1}m_e c^2
\left({m_p\over m_e}{1-\chi_1\over 1+\chi_1}+1\right)\, ,
\end{equation}
where $m_p$ and $m_e$ are the proton and electron masses,
respectively. The latter assumption means that the shock Lorentz factor
is sufficiently large for the random motions of the
particles of the shocked fluid to be highly relativistic, so
that, everywhere behind the shock transition,
\begin{equation}\label{w2}
p_p=e_p/3\, , \quad p_e=e_e/3\, , \quad w=4p=4e/3\, ,
\end{equation}
where $p_e=p_{e^+}+p_{e^-}$, $p=p_p+p_e$, $e_e=e_{e^+}+e_{e^-}$,
and $e=e_p+e_e$.

In the absence of a detailed shock model, it is necessary to specify how the 
postshock internal energy is divided between the proton and
electron/positron components. We do this by parameterizing the ratio of the 
pressure of the electrons and positrons to that
of the total thermal pressure just behind the shock transition,
\begin{equation}\label{p_ep2}
\eta \equiv {p_{e2}\over p_2}\, .
\end{equation}
In view of the assumed postshock equation of state
(eq. [\ref{w2}]), $\eta$ is just the value of the parameter
$\epsilon_e$ (see \S 1) immediately behind the
shock transition ($\eta=\epsilon_{e2}$).

Under the assumption of ideal MHD, the magnetic flux is frozen into the
fluid and remains constant in the shock frame.\footnote{We check on
the validity of this assumption at the end of \S
\ref{hydro}.} For a purely transverse field, this implies
\begin{equation}\label{B}
u_1B_1=u_2B_2=uB\, .
\end{equation}
The momentum and energy 
conservation relations across the shock transition take the form
\begin{eqnarray}\label{momentum_cons}
u_1^2 w_{\rm tot1} + p_{\rm tot1} &=& u_2^2 w_{\rm tot2}+p_{\rm tot2}\, ,\\
\gamma_1 u_1 w_{\rm tot1}&=&\gamma_2 u_2 w_{\rm tot2}\label{energy_cons}\, .
\end{eqnarray}
Since our main interest is in relativistic shock waves, we henceforth set
$\beta_1\approx 1$ and $u_1\approx\gamma_1\gg 1$ and neglect the
second term on the left-hand side of equation (\ref{momentum_cons}). 
Substitution of equations
(\ref{w1}) and (\ref{w2}) into equations
(\ref{momentum_cons}) and (\ref{energy_cons}) then yields, after 
some algebra, an equation for $\beta_2$,
\begin{equation}\label{eq_beta_2}
\beta_2^3-{5u_{A1}^2+4\over 3(u_{A1}^2+1)}
\beta_2^2+{1\over 3}\beta_2+{u_{A1}^2\over 3(u_{A1}^2+1)}=0\, ,
\end{equation}
and an equation for $p_2$,
\begin{equation}\label{p_2}
p_2={u_1^2 u_{A1}^2\rho_1 c^2\over 2\delta_2 u_2^2}\ ,
\end{equation}
where 
\begin{equation}\label{delta_2}
\delta_2\equiv{B_2^2\over 8\pi p_2}={4u_2^2+1\over 2u_2^2 u_{A1}^{-2}-1}
\end{equation}
is the ratio of the magnetic pressure to the thermal pressure just
behind the shock transition, and $u_{A}^2=B^2/4\pi w$ is the
square of the Alfv\'en proper speed (so $u_{A1}^2=B_1^2/4\pi \rho_1 c^2$).
After the gas enters the radiative zone it cools and gets
compressed, resulting in an increase in the magnetic-to-thermal
pressure ratio. In fact, as we show in \S \ref{results},
$\delta(x)$ is a monotonically increasing function of $x$. 

\begin{figure}
\begin{center}
  \plotone{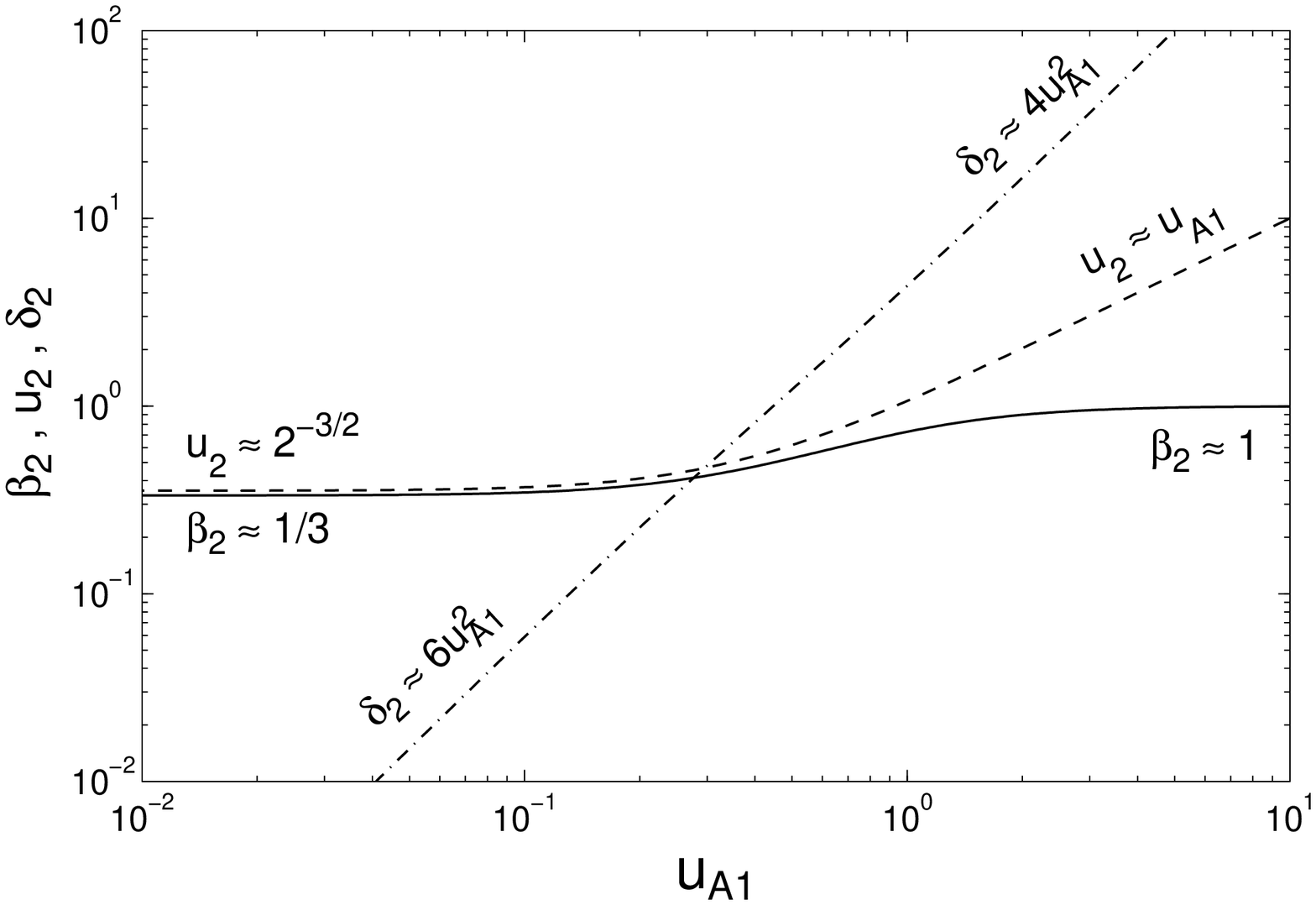} \figcaption[]{The speed ($\beta_2$,
  {\it solid}\/ line), proper speed ($u_2$, {\it dashed}\/
  line), and magnetic-to-thermal
    pressure ratio ($\delta_2$, {\it dash-dotted}\/ line) immediately behind
    the shock transition, as a function of the proper Alfv\'en speed
    $u_{A1}$.  The asymptotic values for $u_{A1}$ much larger or much
    smaller than 1 are given in the figure. In order for a shock wave
    to exist, $u_{A1}$ must be smaller than the upstream speed $u_1$.
\label{beta_u_delta_2}}
\end{center}
\end{figure}

The solution of equation (\ref{eq_beta_2}) is
\begin{equation}\label{beta_2}
\beta_2={1+2u_{A1}^2+\sqrt{16u_{A1}^2(u_{A1}^2+1)+1}\over6(u_{A1}^2+1)}\  .
\end{equation}
In the limit $u_{A1}\ll 1$ the magnetic effects near the shock
transition are weak, and one recovers from equations (\ref{beta_2})
and (\ref{p_2}) the expressions for a nonmagnetic, ultrarelativistic
($\beta_1 \rightarrow 1$) shock, $\beta_2\approx 1/3$ and $p_2\approx
(2/3)u_1^2\rho_1 c^2$. As $u_{A1}$ is increased, $\beta_2$ becomes
larger while $p_2$ becomes smaller. The dependence of $\beta_2$,
$u_2$, and $\delta_2$ on $u_{A1}$ is shown in Figure
\ref{beta_u_delta_2}. Note that, for a given shock speed, $u_{A1}$ has
an upper bound, namely $u_1$. This follows from the fact that, in
order for a shock wave to exist, the relativistic Alfv\'en Mach
number, given by ${M}_{A1}\equiv u_1/u_{A1}$ (e.g., \cite{AK80}), must
exceed 1.

\section{The Radiative Zone}
\label{radzone}

\subsection{Radiation}
\label{radiation}

We take the radiation mechanism behind the shock to be synchrotron
emission and inverse-Compton scattering (IC hereafter) of seed photons
by the relativistic electron-positron component. The seed photons may
consist of internally emitted synchrotron photons, giving rise
to synchrotron self-Compton radiation (SSC), or of an externally
produced population of soft photons, in which case the IC
emission is referred to as external-radiation Compton (ERC).
Although we treat the postshock protons as a relativistic fluid, we
neglect their (much weaker) nonthermal emission.\footnote{If both the
proton and electron components are effectively monoenergetic, then
the ratio of their nonthermal (synchrotron or Thomson-scattering)
cooling times behind the shock is $t_{{\rm
rad},p}/t_{{\rm rad},e} =
[\eta/(1-\eta)][(1-\chi_2)/(1+\chi_2)](m_p/m_e)^4$, which is typically
$\gg 1$. For a discussion of possible observable consequences in GRB
sources of nonthermal emission by protons see, e.g., Zhang \&
M\'esz\'aros (2001) and references therein.}
Furthermore, we neglect any possible transfer of energy between the proton
and electron/positron components in the radiative zone. Under
these assumptions, the protons evolve adiabatically, and the
fraction of the energy dissipated in the shock transition that can
be radiated away is determined uniquely by the parameter $\eta$
defined in equation (\ref{p_ep2}).

The radiating particles are assumed to have a monoenergetic energy
distribution that can be characterized by a random (or
``thermal'') Lorentz factor
$\gamma_e$ (the same for electrons and positrons) that is a function
of the distance $x$ behind the shock.  This approximation to the local
energy distribution is appropriate for electrons and positrons that
have undergone significant cooling. We note in this connection
that observations and particle acceleration models of
relativistic shock sources suggest that the energy distribution immediately
behind the shock transition has the form of a power law. In this
case the radiative zone is expected to consist of a thin layer
just behind the shock transition where most of
the electrons and positrons have not yet cooled significantly
(and are thus not
well approximated by a monoenergetic distribution), followed by a
much wider layer where substantial cooling has already occurred and
where the monoenergetic representation is adequate (see
\cite{GPS2000}). Although the region where the monoenergetic
approximation does not apply is relevant to the spectral
modeling of the source, it can be safely neglected in a
calculation of the cooling-zone dynamics.

The synchrotron emissivity (energy per unit volume per unit time
measured in the rest frame of the fluid) from a population of relativistic
electrons and positrons of total number density $n_e$ that have an
isotropic velocity distribution is given by $\Lambda_{syn}=\sigma_T c
n_e\gamma_e^2B^2/6\pi$ (e.g., \cite{R&L}), where $\sigma_T$ is the
Thomson cross section. To obtain the total emissivity, one
must add the IC component. The IC power of a single electron is
$(4/3)\sigma_T c n_e \gamma_e^2 e_{\rm ph}$, where $e_{\rm ph}$ is the
energy density of photons, assuming an isotropic photon distribution.
This is also the average IC power of an electron for an isotropic
distribution of electron velocities, even when the photon distribution
is not isotropic. Therefore, the total emissivity is
\begin{equation}\label{Lambda}
\Lambda={4\over3}\sigma_T c n_e \gamma_e^2 \left(
e_B +e_{\rm ph}^{\rm syn}+e_{\rm ph}^{\rm ext}\right)\, ,
\end{equation}
where $e_{\rm ph}^{\rm syn}$ and $e_{\rm ph}^{\rm ext}$ are,
respectively, the energy densities of the synchrotron and the
external photons in the local frame. 

We assume that the Thomson optical depth is much smaller than 1, a
situation that often applies in astrophysical relativistic-jet
sources. We further assume that the emitting region is optically thin
to synchrotron self-absorption near the characteristic synchrotron
frequency, which is typical in GRBs and their afterglows. Under these
conditions, most of the radiated energy can freely escape the
system.\footnote{If these conditions are not satisfied, then the
effective value of $t_{\rm rad}$ is increased by opacity effects and
the shock may not even be radiative. In any case, if the
self-absorption frequency exceeds the characteristic frequency, the
shock will become spectrally indistinguishable from a nonradiative
shock (e.g., Granot et al. 2000).}  The average fractional energy
gain, in a single electron scattering, of seed photons that propagate
through the radiative zone is measured by the Compton $y$ parameter.
In the Thomson limit ($\gamma_e h\nu\ll m_e c^2$, where $\nu$ is the
incident photon frequency in the local frame, i.e. the fluid rest
frame), the average fractional energy gain of a seed photon moving at
an angle $\theta=\cos^{-1}\mu$ (measured in the shock frame) with
respect to the $x$ direction is give by
\begin{equation}\label{y_mu}
y(\mu)={4\over 3}\int d\tau_T\gamma_e^2=
{4\over 3}{\sigma_T\over\mu}\int dx\gamma_e^2 n_e\gamma(1-\beta\mu)\, ,
\end{equation}
where $d\tau_T=n_e\sigma_Tds$ is the differential optical depth for
Compton scattering and $ds=\gamma(1-\beta\mu)dx/\mu$ is the
differential of the distance traveled by the photon in the local
frame. One can obtain $y$ by averaging the expression (\ref{y_mu})
over the angular distribution of the seed photons in the shock frame.
In the case of an infinite planar shock, the term $1/\mu$ that appears
in equation (\ref{y_mu}) will cause the integral to diverge
logarithmically for any angular distribution that does not vanish at
$\mu=0$. However, under more realistic circumstances, one may expect
the derived value of $y$ to remain finite. We approximate the integral
in equation (\ref{y_mu}) by setting $\mu = 1$ and then multiplying the
value of the integrand just behind the shock transition by the
characteristic size $x_c$ of the radiative zone (see eq. [\ref{x_c}]
below). This leads to
\begin{equation}\label{y}
y={4\over 3}\sigma_T\gamma_{e2}^2 n_{e2}\gamma_2(1-\beta_2)k_1 x_c\, ,
\end{equation}
where we have introduced a numerical factor $k_1$ (expected to be of
order unity) that incorporates the various uncertainties
involved in our approximation. For $y\ll 1$ only single
scatterings are important, whereas for $y\gtrsim 1$ multiple
scatterings can significantly affect the energy and spectrum of the
scattered photons. However, a typical seed photon with initial
energy $h\nu$ is upscattered on the average to an energy
$\sim\gamma_e^2 h\nu$ in a
single scattering. Usually, the typical electron Lorentz factor
$\gamma_e$ is sufficiently high for the Thomson limit to no longer
apply even in the second scattering ($\gamma_e^3 h\nu>m_e c^2$). This
causes the energy gain in each successive scattering to be
significantly reduced on account of the Klein-Nishina decrease in the
cross section and of electron recoil. We therefore neglect multiple
scatterings in our analysis.

The energy density $e_{\rm ph}^{\rm syn}$ of the synchrotron photons may
be calculated in a manner analogous to that of the Compton $y$
parameter and is subject to similar uncertainties. We treat it
in the same manner as above by introducing a new parameter of order
unity, $k_2$:
\begin{equation}\label{e_syn}
e_{\rm ph}^{\rm syn}=\Lambda_{\rm syn}{x'_c\over c}=
{4\over 3}\sigma_T n_{e2}\gamma_{e2}^2 \gamma_2(1-\beta_2)k_2 x_c=
{k_2\over k_1}y\left({B_2^2\over 8\pi}\right)\, ,
\end{equation}
where $x'_c=x_c\gamma(1-\beta)$ is the length of the trajectory of a
photon in the $x$ direction ($\theta=0$) within the radiative zone, as
measured in the comoving frame. We also consider an external photon
distribution that is isotropic and has an energy density $e_{\rm
ph1}^{\rm ext}$ in the frame of the preshock fluid.  In the local
rest frame of the shocked fluid, the energy density of the external
photons is given by
\begin{equation}\label{e_ext}
{e_{\rm ph}^{\rm ext}\over e_{\rm ph1}^{\rm ext}}=
\gamma_r^2\left(1+{\beta_r^2\over 3}\right)\approx 
{4\over 3}\gamma_1^2\gamma^2(1-\beta)^2\approx
{4\over 3}\gamma_1^2\gamma_2^2(1-\beta_2)^2
\end{equation}
(e.g., \cite{R&L}), where
$\gamma_r=\gamma_1\gamma(1-\beta_1\beta)\approx\gamma_1\gamma(1-\beta)$
is the relative Lorentz factor between the shocked and unshocked
fluids.

The cooling length $x_c$ is the distance behind the shock transition
over which the electrons/positrons lose the bulk of their
internal energy; it is thus the characteristic size of the
radiative zone. Since a significant fraction of
the internal energy must be radiated away before the hydrodynamic
quantities start to deviate significantly from their values just behind
the shock transition, a substantial portion of the radiation
would be emitted under conditions that prevail at the upstream
end of the radiative zone. One can therefore evaluate $x_c$ by
using the parameter values at $x_2$,
\begin{equation}\label{x_c}
x_c\equiv{u_2 c e_{e2}\over \Lambda_2}={2\pi m_e^2 c^4 u_2 n_{e2}\over
(1+b+{k}y)\sigma_T B_2^2 p_{e2}}={16\pi^2m_e^2 c^4 n_{e1}\over
(1+b+{k}y)\sigma_T B_1^4}{u_2^4\over u_1^3}{\phi\delta_2\over\eta}\, ,
\end{equation}
where $k\equiv k_2/k_1$ and $b\equiv e_{\rm ph2}^{\rm
ext}/e_{B2}$,\footnote{Note that the parameter $b$ is also
approximately equal to $e_{\rm ph1}^{\rm
ext}/e_{B1}$, as both energy densities increase by a factor
$\sim \gamma_1^2$ across the shock transition (see
eqs. [\ref{B}] and [\ref{e_ext}]).} and
where we have used equations (\ref{Lambda}) and (\ref{e_syn}).  
From equations 
(\ref{x_c}) and (\ref{y}) we obtain an equation for $y$:
\begin{equation}\label{y2} 
(k y)^2+(1+b)ky=k_2u_2\gamma_2 (1-\beta_2){e_{e2}\over e_{B2}}=
k_2u_2\gamma_2(1-\beta_2){3\eta\over \delta_2}\equiv a\, ,
\end{equation}
which yields
\begin{equation}\label{y3}
ky={(1+b)\over 2}\left({\sqrt{1+{4a\over(1+b)^{2}}}\,-1}\right)\approx 
\left\{\matrix{\sqrt{a} & \ \ 1,\, b^2\ll a \cr & \cr 
a & \ \ a,\, b\ll 1 \cr & \cr a/b & \ \ 1,\, a\ll b^2}\right. \, 
\end{equation}
In the first two limits the external photons are not important, and we
obtain the familiar expressions for emission dominated by SSC and
synchrotron, respectively (\cite{SNP}).  The physical meaning of each
of these limits can be understood by comparing the relative magnitudes
of the different components of the energy density, keeping in mind
that $a\approx e_{e2}/e_{B2}$, $b=e_{\rm ph2}^{\rm ext}/e_{B2}$, and
$e_{\rm ph2}^{\rm syn}\approx ye_{B2}$. In the first limit $y\approx
a^{1/2}\gg 1$ and $b^2\ll a$, implying that $e_{B2},\, e_{\rm
  ph2}^{\rm ext}\ll e_{\rm ph2}^{\rm syn}$, which according to
equation (\ref{Lambda}) means that most of the radiated power goes
into SSC.  One also obtains $e_{e2}\approx y e_{\rm ph2}^{\rm syn}$,
which is consistent with this interpretation, since, in a steady-state
radiative shock, the incoming flux of internal energy
($\beta_2ce_{e2}\approx ce_{e2}$) equals the outgoing radiation flux,
which in this limit is dominated by SSC and is therefore approximately
equal to $cye_{\rm ph2}^{\rm syn}$.  In the second limit $y\approx
a\ll 1$ and $b\ll 1$, implying that $e_{\rm ph2}^{\rm ext}, \, e_{\rm
  ph2}^{\rm syn}\ll e_{B2}$, which means that synchrotron radiation is
dominant over IC. In this case we also have $e_{e2}\approx e_{\rm
  ph2}^{\rm syn}$, which implies that all the internal energy goes into
synchrotron radiation. In the third limit $y\approx a/b$ and $1,a\ll
b^2$, implying $e_{\rm ph2}^{\rm syn},\, e_{B2}\ll e_{\rm ph2}^{\rm
  ext}$, which means that most of the emitted power goes into
IC-scattered external photons (ERC). This conclusion is consistent
with the fact that $e_{e2}\approx ye_{\rm ph2}^{\rm ext}$ in this case,
which means that all the internal energy is going into ERC. Note that
the value of $y$ ceases to be pertinent in this limit as it no longer
gives the ratio of the IC and synchrotron emissivities (in particular,
$y$ could be $\ll 1$ even as IC remains dominant).

The photon energy densities $e_{\rm ph}^{\rm ext}$ and $e_{\rm
ph}^{\rm syn}$ are approximately constant throughout the radiative zone,
whereas, as we demonstrate in \S \ref{results}, the magnetic energy
density $e_{B} \propto B^2$ may be considerably amplified if $\eta\approx 1$.
Thus, even if synchrotron emission is not dominant immediately
behind the shock transition (as in limits 1 and 3 above), it may become so
further downstream. However, by the time this happens,
only a small fraction of the initial internal energy would
typically be left in the electron/positron component. Therefore,
the total emission would still be dominated by IC (either SSC, as in the first
limit, or ERC, as in the third limit). Nevertheless, given that
the characteristic IC frequency is usually much higher than that
of the synchrotron emission (by a factor $\sim \gamma_e^2$ in
the case of SSC), it is likely even in this case that only the synchrotron
component is detected when the observation frequency is not
too high.

\subsection{Hydrodynamics}
\label{hydro}
 
The steady-state momentum and energy conservation equations in
the shock frame can be written as
\begin{eqnarray}\label{momentum_evol}
\partial_x(u^2 w_{\rm tot}+p_{\rm tot})=
-u\Lambda/c\, ,\\
\partial_x(\gamma u w_{\rm tot})=
-\gamma\Lambda/c \label{energy_evol}\, .
\end{eqnarray}
The terms on the right-hand sides of these equations represent
the contribution of the radiation field to the momentum and
energy fluxes, respectively. Their forms are appropriate when
the emitted or scattered radiation exerts no net force on the fluid;
in particular, they apply when the radiation field is isotropic in the fluid
frame.\footnote{Note that the corresponding terms in equations (32)
and (33) of Cohen et al. (1998) evidently need to be corrected.}  If
the magnetic field is ordered, as we have assumed, then the
synchrotron emission of any given particle is not isotropic in the
fluid rest frame; however, so long as the electron/positron rest-frame
velocity distribution is isotropic, the total synchrotron radiation
will carry no net momentum flux in this frame and will, therefore,
exert no net force on the fluid. The resulting synchrotron
photon energy density (eq.  [\ref{e_syn}]), when scattered by the
isotropic electron/positron distribution, gives rise to an SSC emissivity
that is isotropic in the fluid frame. 
The same applies to an ERC component that arises from the scattering
(by electrons/positrons that are isotropic in the local rest frame)
of an isotropic (in the rest frame of the central source) external
radiation field (where now the photon energy density that appears in
the isotropic-emission expression, in the local rest frame, is given
by eq.  [\ref{e_ext}]; see \cite{D&S}).\footnote{We ignore in our
  discussion the effect of the radiation drag force (corresponding to
  a nonvanishing momentum flux of intercepted photons in the fluid
  rest frame) that is exerted by an external radiation field on the
  bulk motion of the fluid. This is consistent with our neglect of any
  variations in the gross properties of the shock that occur on time
  scales $\gtrsim t_{\rm dyn}$.}

From equations (\ref{momentum_evol}) and (\ref{energy_evol}) one
obtains an equation for the entropy:
\begin{equation}\label{entropy}
w_{\rm tot} \partial_x u + u\partial_x
(w_{\rm tot}-p_{\rm tot})=-\Lambda/c\, .
\end{equation}
The magnetic field evolves according to equation (\ref{B}) and the protons 
evolve adiabatically,
\begin{equation}\label{protons}
{p_p\over p_{p2}}=\left({n_p\over n_{p2}}\right)^{4/3}=
\left({u\over u_2}\right)^{-4/3}\ ,
\end{equation}
so the terms that involve them in equation (\ref{entropy}) cancel out. One is
left with
\begin{equation}\label{entropy_e}
4p_e\partial_x u+3u\partial_x p_e=-\Lambda/c\, .
\end{equation}

Eliminating the radiative terms from equations (\ref{momentum_evol}) and
(\ref{energy_evol}) results in
\begin{equation}\label{dp_du}
{dp\over du}={B_2^2u_2^2\over 4\pi u^3(1+u^2)}-{4up\over (1+u^2)}\ ,
\end{equation}
or, in terms of the dimensionless pressure $\tilde{p}\equiv p/p_2$,
\begin{equation}\label{dptd_du}
{d\tilde{p}\over du}={2\delta_2 u_2^2 u^{-3}-4u\tilde{p}\over (1+u^2)}\ .
\end{equation}
Equation (\ref{dptd_du}) can be integrated to give
\begin{equation}\label{p_tilde}
\tilde{p}(u)={(1+u_2^2)^2+\delta_2
\left[1-(u_2/u)^{2}+2u_2^2\ln(u/u_2)\right]\over (1+u^2)^2}\ .
\end{equation}
Using equations (\ref{p_tilde}), (\ref{protons}), and (\ref{p_ep2}), we deduce
\begin{equation}\label{p_e_tilde}
\tilde{p}_e(u)={\tilde{p}-(1-\eta)\tilde{p}_p\over\eta}=\quad
\quad\quad\quad\quad\quad\quad\quad\quad\quad\quad\quad\quad\end{equation}
$$
{(1+u_2^2)^2+\delta_2
\left[1-(u_2/u)^{2}+2u_2^2\ln(u/u_2)\right]\over \eta(1+u^2)^2}
-{1-\eta\over\eta}\left({u_2\over u}\right)^{4/3}\ , 
$$
where $\tilde{p}_p\equiv{p_p/p_{p2}}$ and
$\tilde{p}_e\equiv{p_e/p_{e2}}$.  We already have explicit expressions
for $p_e,\ p_p,\ B,\ n_p,\ n_e$, and $\gamma_e=3p_e/n_em_ec^2$ as
functions of $u$. To obtain their values as a function of the distance
behind the shock, we need to know $u(x)$. Introducing the
dimensionless variable $\xi\equiv x/x_c$ and using equations
(\ref{p_2}), (\ref{Lambda}), (\ref{x_c}), (\ref{entropy_e}), and
(\ref{dp_du}), we get
\begin{equation}\label{du_dxi}
{du\over d\xi}=-{3\eta u_2^2 u(1+u^2)\tilde{p}(u)_e^2
\left[(1+Y)/(1+Y_2)\right]\over 4(1-2u^2)
\left[(1-\eta)u_2^{4/3}u^{2/3}+\eta u^2\tilde{p}(u)_e\right]
+6\delta_2 u_2^2}\ ,
\end{equation}
where the function
\begin{equation}\label{Y} 
Y(u)=(b+ky)\left ( {u \over u_2} \right )^2 
\end{equation}
represents the ratio of the local IC (including SSC and ERC) and synchrotron
emissivities, and $Y_2=(b+ky)$ is the value of $Y$ just behind
the shock transition. Note that the only dependence of the
function $Y(u)$ on the Compton $y$ parameter (see eq. [\ref{y}])
is through the combination $ky$, which, in turn, is given (in
terms of the parameters $a$ and $b$) by equation (\ref{y3}). As
both $u_2$ and $\delta_2$ are functions of $u_{A1}$ (see eqs.
[\ref{delta_2}] and [\ref{beta_2}]), $a$ is determined by the
parameters $k_2$, $u_{A1}$, and $\eta$. For simplicity, we set
$k_2 = 1$ in the rest of this paper.  Therefore, $Y(u)$ and the
right-hand side of equation (\ref{du_dxi}) are
specified by the parameters $\eta$, $u_{A1}$, and $b$, and $u(\xi)$ can
be obtained by solving equation (\ref{du_dxi}) numerically,
using the boundary condition $u(\xi=0)=u_2$.

Before proceeding to present the solutions of the above set of equations, we
digress briefly to check on the validity of the ideal-MHD approximation that
underlies our analysis. Under the assumption of ideal MHD, the electric field 
vanishes in the rest frame of the fluid. The magnetic field in the 
shock frame is then given by $\gamma\vec{B}=u_1B_1\hat{e}_y/\beta(x)$, where 
we have chosen the $y$ axis to point along the direction of the magnetic
field. This implies a current density
\begin{equation}\label{j}
\vec{j}={c\over 4\pi}\nabla\times(\gamma\vec{B})=
-{u_1 B_1 c\over 4\pi\beta^2}{d\beta\over dx}\hat{e}_z\, .
\end{equation}
Since in our case $\vec{j}$ is perpendicular to the
direction of the velocity, it has the same magnitude in the shock
frame. In order for the ideal-MHD approximation to be valid, the
current density in the fluid rest frame cannot exceed $j_{\rm
max}=nec=n_2u_2ec/u(x)$, where $n=n_e+n_p$ is the total number
density of charged particles (e.g., \cite{M&M}).  Furthermore,
so long as $j\ll j_{\rm max}$, the effect of anomalous 
resistivity associated with current-driven plasma instabilities is expected to
be small (e.g., \cite{SDD}). The condition $j<j_{\rm max}$ translates into
\begin{equation}\label{check_1}
{\gamma\over\beta}\left|{d\beta\over dx}\right|<{4\pi e n_2\over B_2}\ .
\end{equation}
In the radiative zone $(\gamma/\beta)|d\beta/dx|\lesssim
\gamma_2/x_c\approx 1/x_c$, so equation (\ref{check_1}) may be written as
\begin{equation}\label{check_2}
1<{4\pi e n_2 x_c\over B_2}={64\pi^3m_e^2c^4e\over\sigma_T}
{n_1 n_{e1}u_2^4\over B_1^5u_1^3}
\left({n_2u_2\over n_1u_1}\right){\phi\delta_2\over\eta(1+Y_2)}\ .
\end{equation}
Note that this inequality can be rewritten in terms of the postshock
electron Larmor radius $r_{\rm L,2}$ as $r_{\rm L,2}<3x_c/2\delta_2$. 
This shows that, if ideal MHD holds in the
radiative zone, then the condition for the fluid approximation to be
valid there (essentially $r_{\rm L,2}\ll x_c$) will also be satisfied.
We now check whether the inequality (\ref{check_2})
holds under the most unfavorable choice of parameters. The latter is
obtained by taking $\eta=\phi=1$ and $\delta_2\ll 1$, which implies
$n_2u_2=n_1u_1$, $u_{A1}\ll 1$, $u_2\approx 1/\sqrt{8}$, and
$\delta_2\approx 3B_1^2/2\pi\rho_1 c^2$, resulting in
\begin{equation}\label{check_3}
1<{3\pi^2m_e^2c^2e\over 2\sigma_T(1+Y_2)}{n_{e1}n_1\over\rho_1 B_1^3 u_1^3}\ .
\end{equation}
The minimal value of $n_{e1}/\rho_1$ is $1/m_p$, which is obtained
for $\chi_1=0$. This leads to
\begin{equation}\label{check_4}
1<{3\pi^2m_e^2c^2e\over 2\sigma_T m_p(1+Y_2)}{n_1\over B_1^3 u_1^3}
={9.5\times 10^6 n_{p1,0}\over(1+Y_2)B^3_{1,0} u_1^3}
\approx {2.2\times 10^8 n_{p1,0}\over(1+Y_2)B^3_{2,0}}\ ,
\end{equation}
where here and below numerical subscripts (such as the subscript
0 in $n_{p1,0}$) refer to powers of 10 in c.g.s units.

As an illustration we apply this condition to GRBs, considering both
the prompt high-energy emission (resulting from internal shocks within
the flow) and the lower-energy afterglow (arising from the interaction
of the flow with the surrounding medium). We begin with internal
shocks. We use the conventional parameterization
$B_2^2=8\pi\epsilon_{B2} e_2$ (where we identify
$\epsilon_{B2}=\delta_2/3$), and write (e.g., Piran 1999)
$e_2=(\gamma_{12}-1)(4\gamma_{12}+3)n_{p1} m_p c^2$, $n_{p1}=E/4\pi
m_p R^2\gamma^2c^3t_{\rm dur}$, and $R=2\gamma^2ct_{\rm dur}\zeta$,
where $E$ is the total energy deposited in the flow, $\gamma$ is the
typical Lorentz factor of the flow with respect to the central object
(and the observer), $\gamma_{12}$ is the relative Lorentz factor
between the shocked and the unshocked fluids, $t_{\rm dur}$ is the
duration of the burst, and $\zeta=\delta t/t_{\rm dur}$ is a measure
of the variability of the burst, with $\delta t$ being the smallest
time scale for significant variation during the burst. Thus we obtain
\begin{equation}\label{check_IS}
1<{1.3\times 10^5 t_{\rm
dur,1}^{3/2}\gamma_2^3\zeta_{-2}\over(1+Y_2)\epsilon_{B2}^{3/2}E_{52}^{1/2}
\left[(\gamma_{12}-1)(4\gamma_{12}+3)\right]^{3/2}}\approx
{3.6\times 10^3 t_{\rm dur,1}^{3/2}\gamma_2^3\zeta_{-2}\over(1+Y_2)
\epsilon_{B2}^{3/2}E_{52}^{1/2}}\ ,
\end{equation}
where we used $\gamma_{12}=2$ on the right-hand side (consistent with
the expectation that $\gamma_{12}$ is of the order of 1). Turning now
to the afterglow stage of GRBs, we evaluate the inequality
(\ref{check_3}) by using $e_2=4\gamma^2n_{p1}m_pc^2$ in the
parametrized expression for $B_2$. This yields
\begin{equation}\label{check_ES}
1<1.2 \times 10^2\,
(1+Y_2)^{-1}\epsilon_{B2}^{-3/2}n_{p1,3}^{-1/2}\gamma_2^{-3}\, ,
\end{equation}
where we scaled $\gamma$ and $n_{p1}$ by their rough upper
limits for typical afterglows. (The Lorentz factor $\gamma$
decreases during the afterglow evolution, as does $n_{p1}$ if
the shock propagates into a stellar-wind environment.) Since the factor
$(1+Y_2)\epsilon_{B2}^{3/2}$ in equations (\ref{check_IS}) and
(\ref{check_ES}) never exceeds 1 in the absence of a strong
external radiation field (being approximately equal to ${\rm
max}\{\epsilon_{B2}^{3/2},\epsilon_{B2}\eta^{1/2}\}$; see
eq. [\ref{y3}]), we conclude that the use of the
MHD approximation for studying the cooling layers behind
radiative shocks should be well justified throughout the 
evolution of typical GRBs and their afterglows.

\subsection{Results}
\label{results}

\begin{figure}
\begin{center}
  \plotone{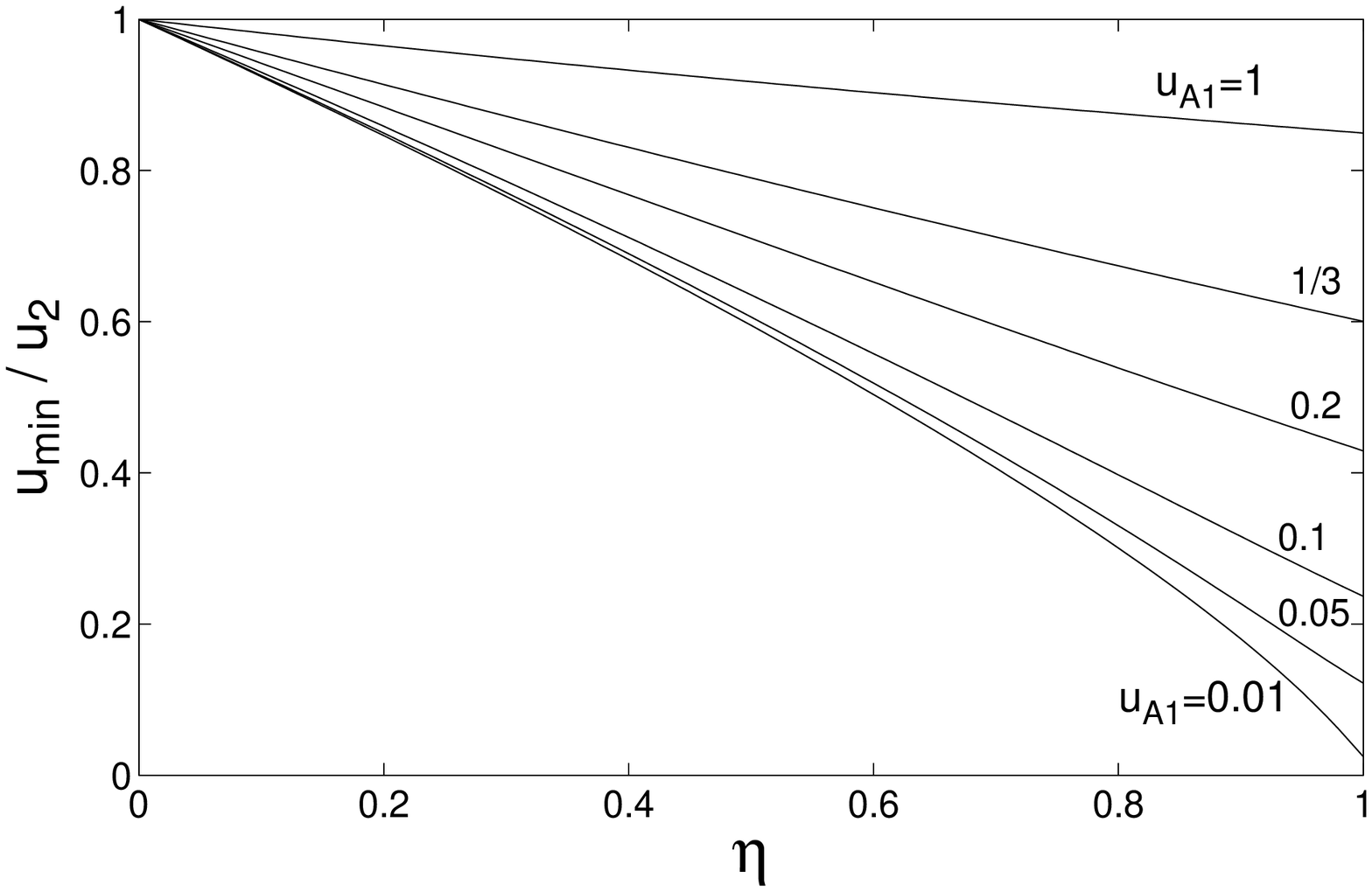} \figcaption[]{The asymptotic proper
    speed $u_{\rm min}$ (in units of $u_2$) as a function of $\eta$ for
    several values of $u_{A1}$. 
\label{u_min_eta}}
\end{center}
\end{figure}

\begin{figure}
\begin{center}
  \plotone{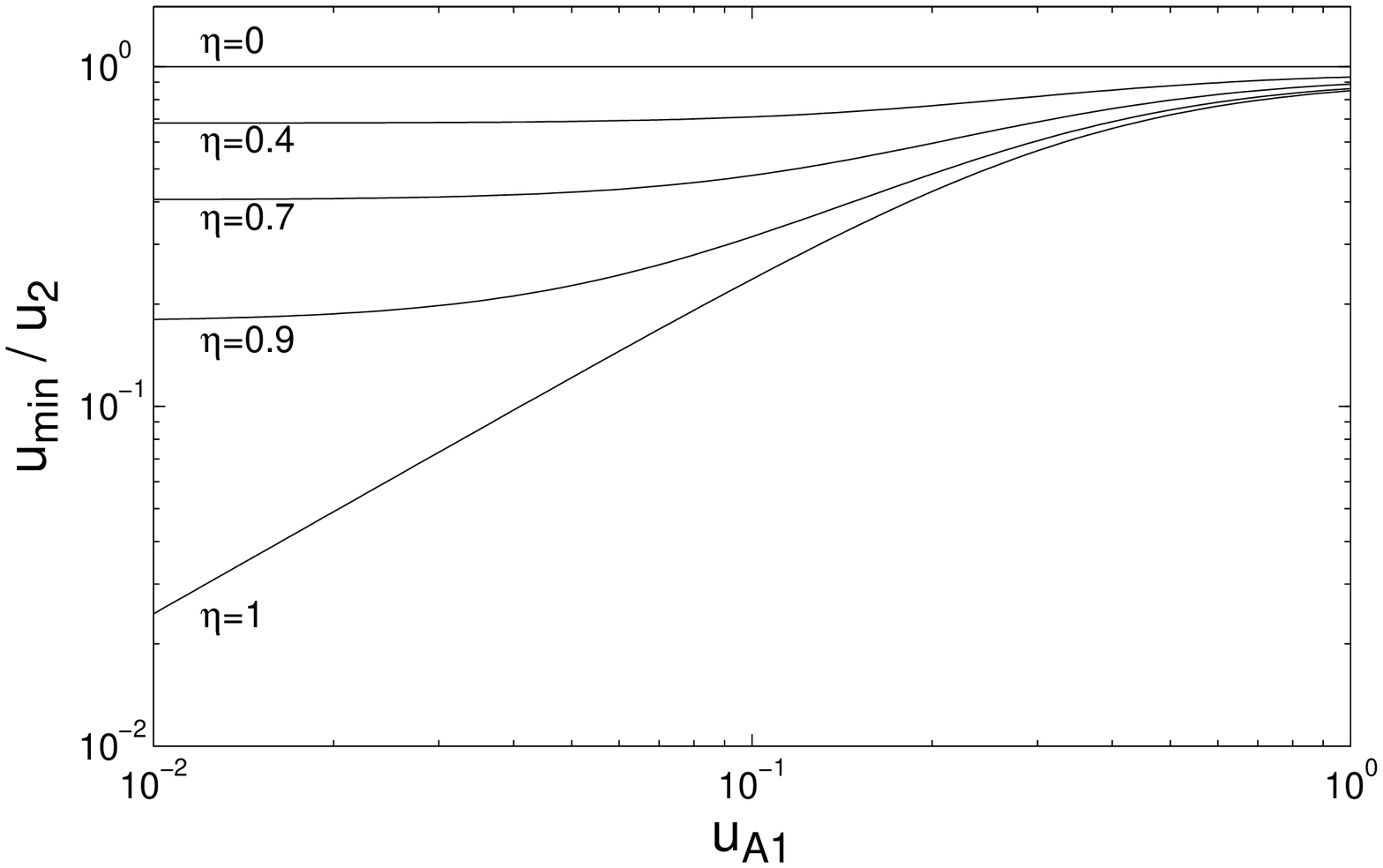} \figcaption[]{The asymptotic proper
    speed $u_{\rm min}$ (in units of $u_2$) as a function of $u_{A1}$ for
    several values of $\eta$.
\label{u_min_M}}
\end{center}
\end{figure}

At large distances behind the shock ($\xi\gg 1$) the electrons and
positrons are left with only a small fraction of their internal
energy, and all hydrodynamic quantities approach some constant
asymptotic value: $u\approx u_{\rm min}$, $B\approx B_{\rm
  max}=B_2u_2/u_{\rm min}$. By equating the radiative cooling time
with the flow time, one infers that $\gamma_e\approx\gamma_{e2}/\xi$
in that region (e.g., Granot et al. 2000). Also, $n_e\approx
n_{e2}u_2/u_{\rm min}$, and hence $\tilde{p}_e\approx u_2/u_{\rm
  min}\xi$ for $\xi\gg 1$. The fact that $\tilde{p}_e(\xi)>0$ for any
finite value of $\xi$ implies that $u(\xi)>u_{\rm min}$, so $u$
becomes equal to $u_{\rm min}$ only asymptotically. One can determine
$u_{\rm min}$ by setting equation (\ref{p_e_tilde}) equal to zero and
(numerically) solving for $u$: this yields $u_{\rm min}$ as a function
of $\eta$ and $u_{A1}$.  In Figures \ref{u_min_eta} and \ref{u_min_M}
we fix one of these variables and show $u_{\rm min}$, normalized by
$u_2$, as a function of the other variable. In the limit
$\eta\rightarrow 0$ there are no radiation losses, so the speed of the
shocked fluid does not change: $u_{\rm min}=u_2$. In the limit
$\eta\rightarrow 1$ the electrons and positrons contain all the
postshock internal energy and radiate it away, and the attendant
compression eventually renders the magnetic pressure dominant.  Since
the magnetic field is amplified by a factor $\sim u_2/u_{\rm min}$ in
the radiative zone (see eq. [\ref{B}]), a smaller initial magnetic
field (smaller $u_{A1}$) results in a lower value of $u_{\rm
  min}/u_2$. Furthermore, the total pressure, $p_{\rm tot}$, is
approximately constant throughout the radiative zone, and for
$u_{A1}\ll 1$ it becomes independent of $u_{A1}$
($p_{\rm tot}\approx p_2\approx (2/3)u_1^2\rho_1c^2$). Since in the
limit $\eta\rightarrow 1$ the magnetic pressure eventually becomes
dominant ($B_{\rm \max}^2/8\pi\approx p_{\rm tot}$), $B_{\rm \max}$
also becomes independent of $u_{A1}$ for $u_{A1}\ll 1$, and we obtain
$u_{\rm min}/u_2=B_2/B_{\rm \max}\propto u_{A1}$, as can be seen
in Figure \ref{u_min_M}.

\begin{figure}
\begin{center}
\plotone{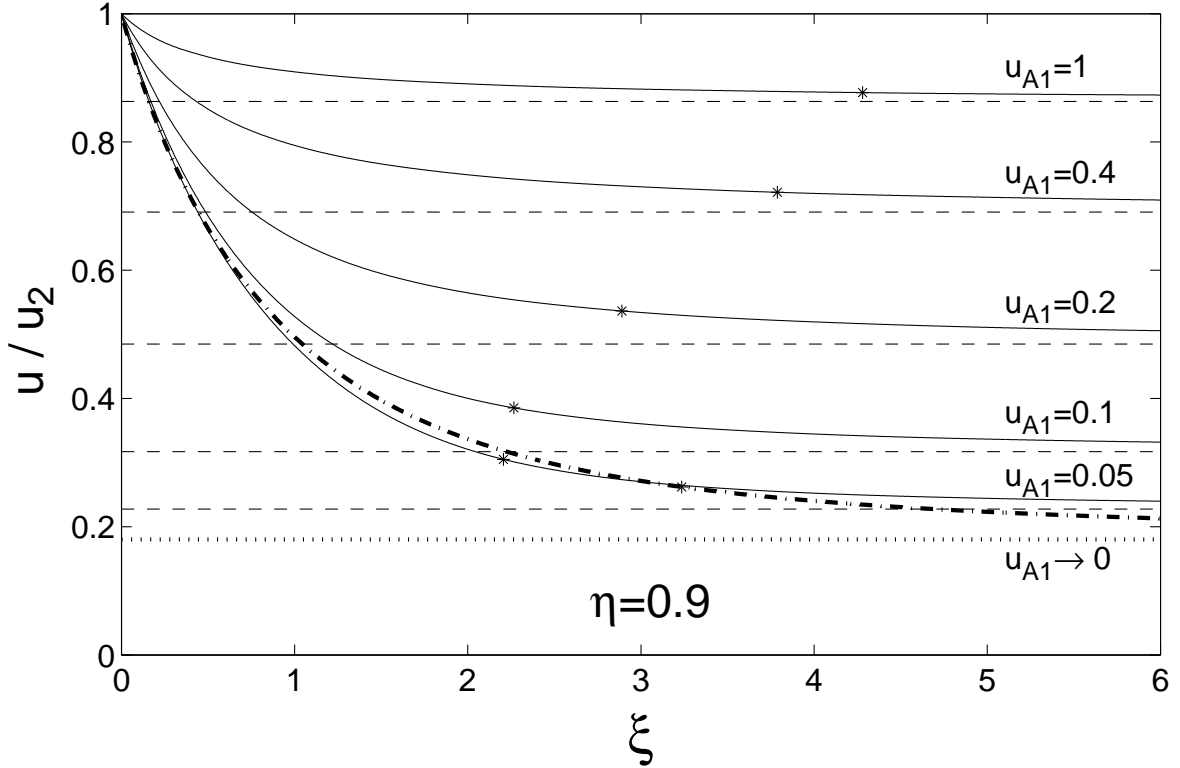} \figcaption[]{The proper speed $u$,
normalized by the postshock value $u_2$, as a function of the
normalized distance $\xi\equiv x/x_c$ for $\eta=0.9$, $b=0$, and
several values of $u_{A1}$ ({\it solid}\/ lines). The {\it dashed}\/ lines
represent the corresponding asymptotic values ($u_{\rm min}$). The {\it
boldface dash-dotted}\/ line depicts the limit $u_{A1}\rightarrow 0$,
and the {\it boldface dotted}\/ line is the corresponding asymptotic
value. An asterisk is placed on the solid lines and on the
dash-dotted line at the point where $u$ has been reduced to 90\% of its
asymptotic value: $(u-u_{\rm min})/(u_2-u_{\rm min})=0.1$. 
\label{u_xi_M}}
\end{center}
\end{figure}

Figure \ref{u_xi_M} displays $u(\xi)$, normalized by $u_2$, for
$\eta=0.9$, $b=0$, and several values of $u_{A1}$. For large values of
$u_{A1}$ ($\gtrsim 0.08$) the asymptotic value $u=u_{\rm min}$ (shown
by a dashed line) is approached at smaller values of $\xi$ (traced by
the asterisks) as $u_{A1}$ decreases, whereas the converse occurs for
smaller values of $u_{A1}$.
This can be understood as follows. For large values of
$u_{A1}$ the initial magnetic field $B_2$ is strong and $Y_2$ is close
to $1$, so synchrotron cooling is important early on. As the internal
energy is radiated, the fluid becomes denser and the magnetic field grows
stronger, which enhances the synchrotron emission. Since the
cooling length is calculated by using the initial synchrotron
emissivity, the value of $x_c$ is overestimated in this case and
hence the asymptotic value of $u=u_2$ is approached at lower values of
$\xi$.  This effect is stronger the larger the value of $B_{\rm
max}$, which for a fixed $\eta$ is obtained for lower
values of $u_{A1}$. However, for sufficiently small values
of $u_{A1}$, $Y_2$ comes to exceed 1 and IC becomes the dominant cooling
mechanism. For $u_{A1}\ll 1$ synchrotron radiation hardly
affects the dynamics, and $x_c$ is determined by the IC
emissivity just behind the shock transition. Since, unlike
the magnetic energy density, the photon energy density that
determines the IC cooling rate does not increase with $\xi$,
the value of $x_c$ better approximates the actual cooling length,
and consequently $u_{\rm min}$ is approached at larger values of
$\xi$ for smaller $u_{A1}$.

\begin{figure}
\begin{center}
  \plotone{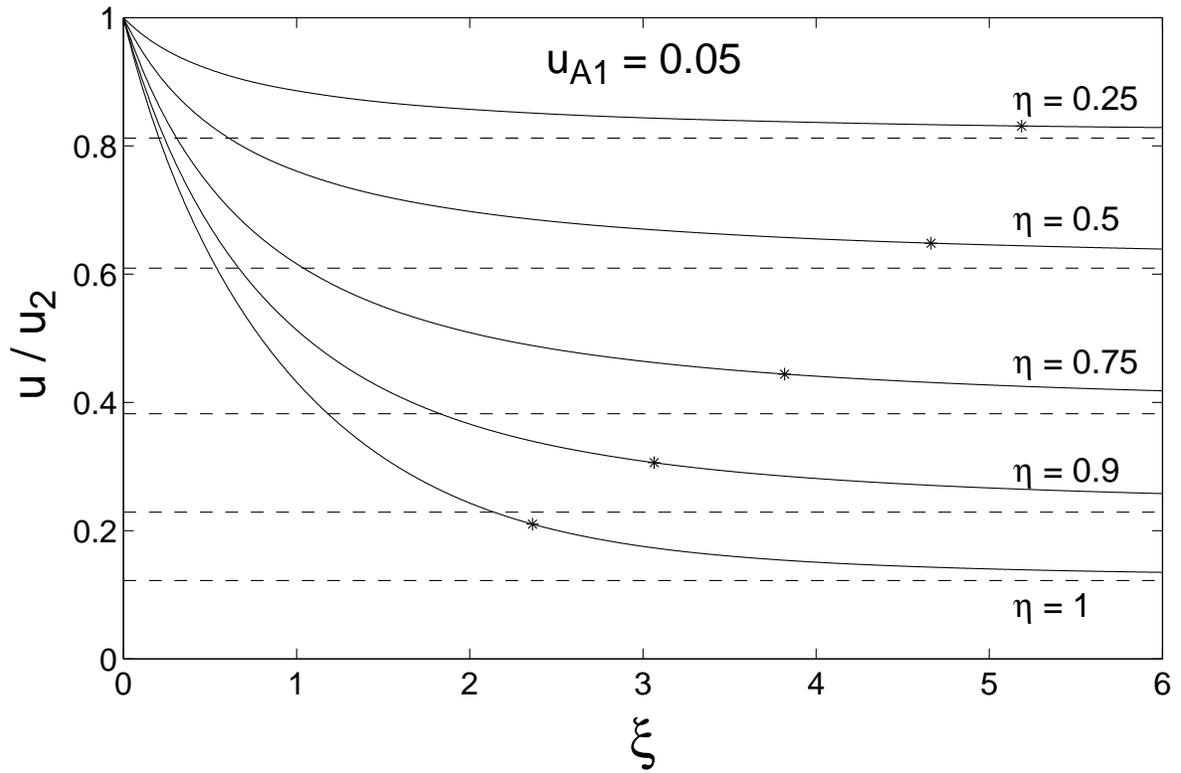} \figcaption[]{The proper speed $u$,
    normalized by the postshock value $u_2$, as a function of the normalized
    distance $\xi\equiv x/x_c$ for $u_{A1}=0.05$, $b=0$, and
    several values of $\eta$ ({\it solid}\/ lines). The {\it dashed}\/ lines
    and the asterisk symbols have the same meaning as in Fig. \ref{u_xi_M}.
\label{u_xi_eta}}
\end{center}
\end{figure}

Figure \ref{u_xi_eta} shows $u(\xi)$, normalized by $u_2$, for
$u_{A1}=0.05$, $b=0$, and several values of $\eta$. In this case the asymptotic
value of $u$ is approached at smaller values of $\xi$ for
larger values of $\eta$, which correspond to larger values of
$B_{\rm max}$ and a stronger enhancement of the synchrotron cooling.

It is straightforward to verify from equations (\ref{p_e_tilde})
and (\ref{du_dxi}) that $p_e$ is a monotonically increasing function
of $u$ and that $u$ is a monotonically decreasing function of $x$.
Therefore, as expected, both $p_e$ and $u$ decrease the
further one gets from the shock front.

\begin{figure}
\begin{center}
  \plotone{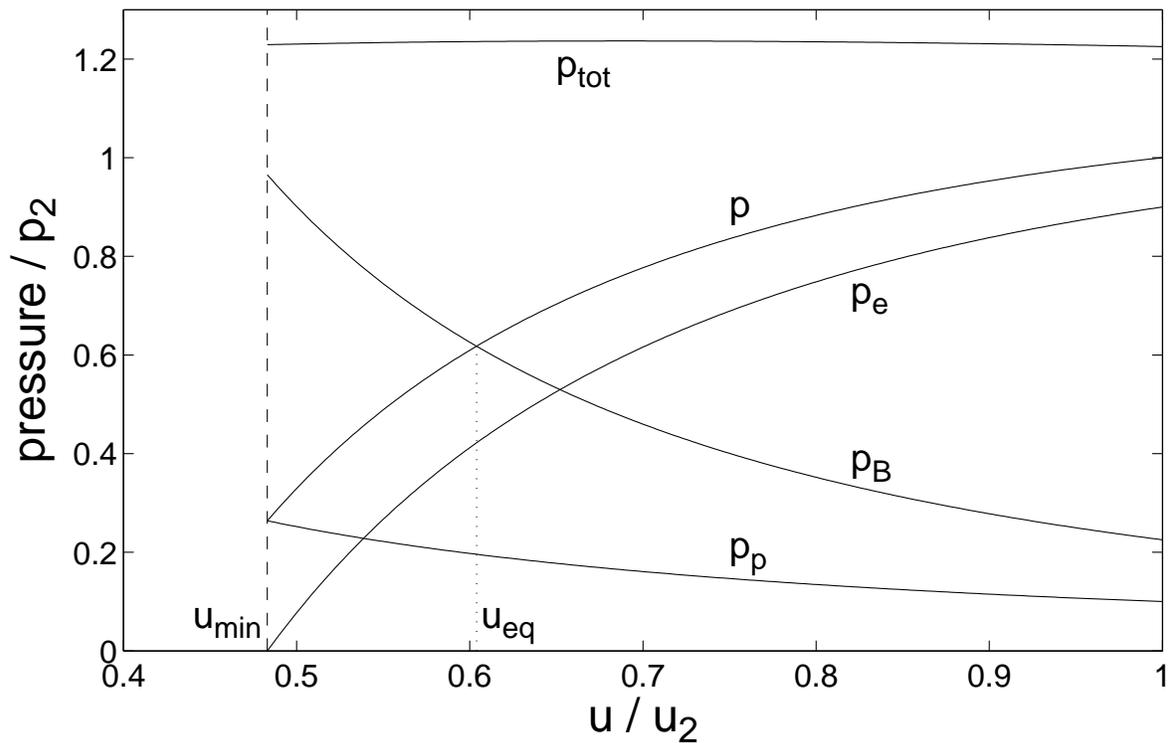} \figcaption[]{The pressures of the
    electron/positron ($p_e$) and proton ($p_p$) components, the total
    thermal pressure ($p=p_e+p_p$), the magnetic field pressure
    ($p_B=B^2/8\pi$) and the total pressure ($p_{\rm tot}=p+p_B$), all
    normalized by the postshock thermal pressure $p_2$, as a function
    of the normalized proper speed $u/u_2$ for $\eta=0.9$ and
    $u_{A1}=0.2$.
\label{p_u}}
\end{center}
\end{figure}

\begin{figure}
\begin{center}
  \plotone{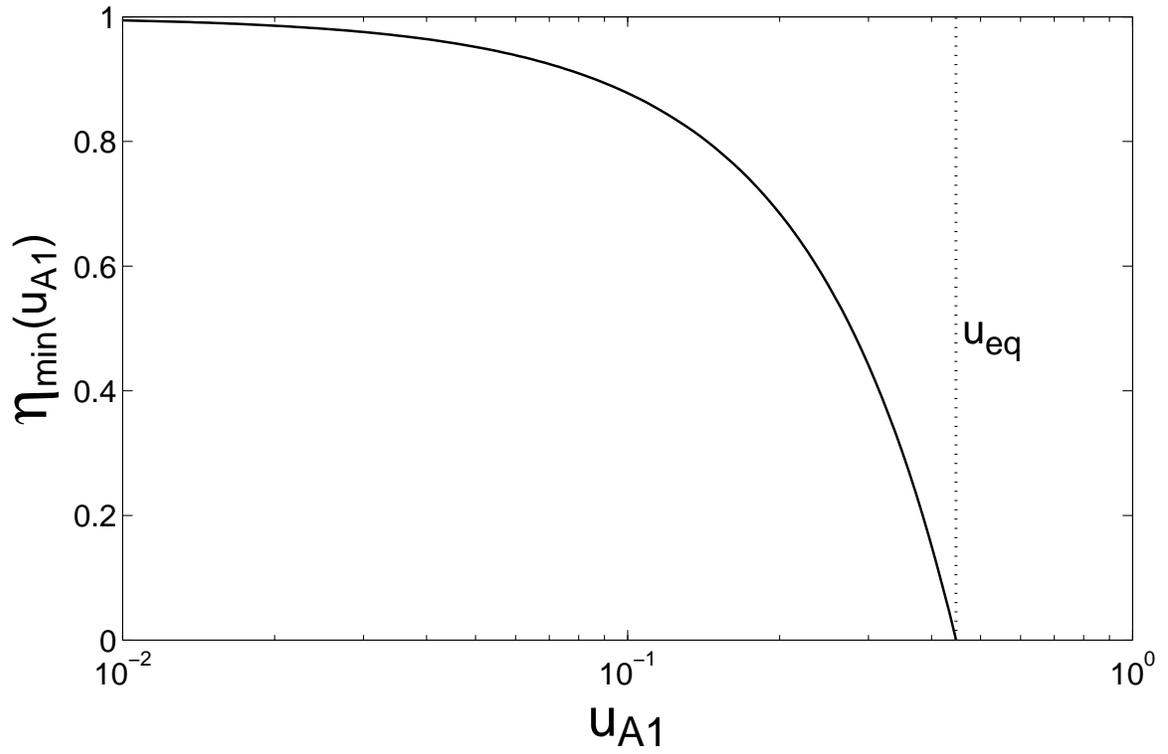} \figcaption[]{The minimum value of
    $\eta$ for which the magnetic field in the radiative zone reaches 
    equipartition with the total thermal pressure, as a
    function of $u_{A1}$. For $u_{A1}>u_{\rm eq}$ the magnetic
    field amplitude just behind the shock transition is already
    above equipartition.
\label{eta_min}}
\end{center}
\end{figure}

\begin{figure}
\begin{center}
  \plotone{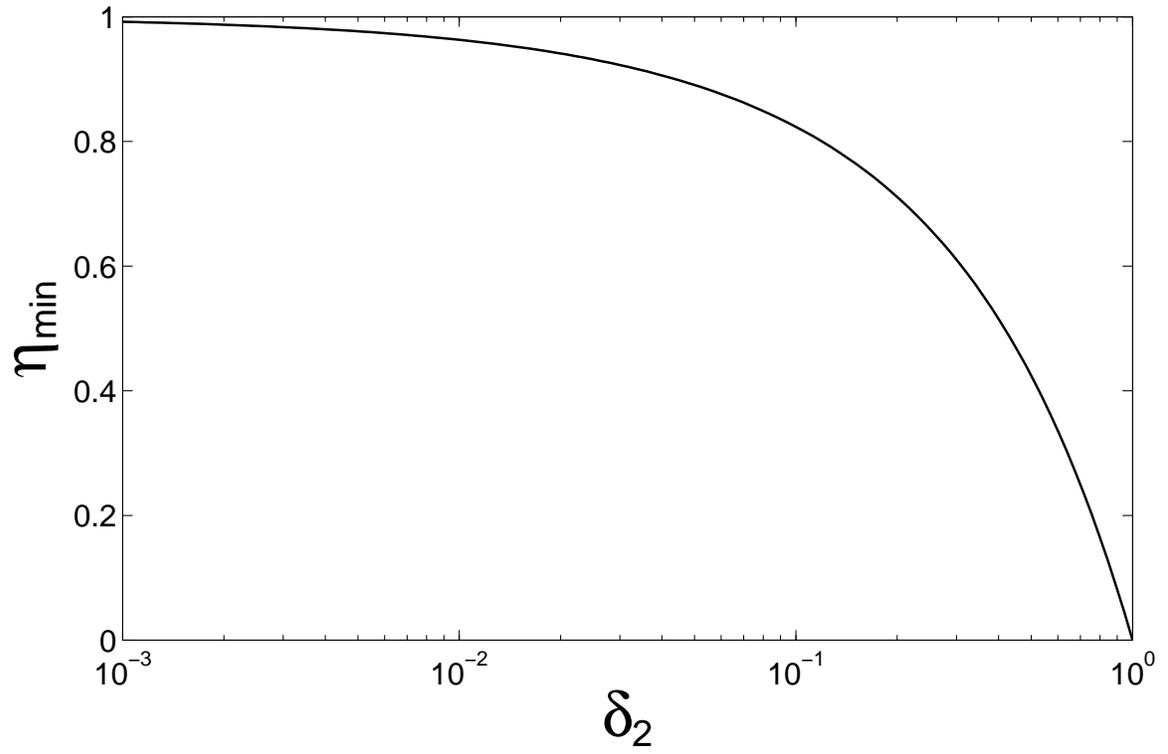} \figcaption[]{The minimum value of
    $\eta$ for which the magnetic field in the radiative zone reaches 
    equipartition with the total
    thermal pressure, as a function of $\delta_2$, the ratio of the 
    postshock magnetic pressure to its local equipartition value.
\label{eta_min_delta2}}
\end{center}
\end{figure}

To demonstrate the evolution of the various components of the
pressure, we show in Figure \ref{p_u} the dependence of $p$, $p_e$, $p_p$, $p_B$
and $p_{\rm tot}$ (each normalized by $p_2$) on $u$ for $\eta=0.9$ and $u_{A1}=0.2$.
Equipartition occurs when the total thermal pressure equals the
magnetic pressure, $p(u_{\rm eq})=p_B(u_{\rm eq})$, with $u_{\rm
eq}$ denoting the value of $u$ at equipartition. However, this
condition is not satisfied in all cases.
There is a critical value of $u_{A1}$, which we denote by $u_{\rm eq}$,
for which $\delta_2(u_{\rm eq})=1$. We find that $u_{\rm eq}=0.447$. For
$u_{A1}>u_{\rm eq}$ the magnetic field just behind the shock
transition is already
above equipartition, $B_2>B_{\rm eq2}$, where $B_{\rm eq}$ is given by
$B_{\rm eq}= (8\pi p)^{1/2}$.  On the other hand, $B_2<B_{\rm eq2}$
for $u_{A1}<u_{\rm eq}$, and only if $\eta$ is above some critical value
$\eta_{\rm min}(u_{A1})$ will the asymptotic magnetic field be
above equipartition.
Since $u_2$, and therefore also $\delta_2$, are functions of $u_{A1}$,
$\tilde{p}$ is a function only of $u_{A1}$ and $u$ (see eq.
[\ref{p_tilde}]). Therefore, $\eta_{\rm min}(u_{A1})$ may be found by
(numerically) solving the equation
\begin{equation}\label{eta_min_u_A1}
\tilde{p}\left[u_{A1},u_{\rm min}(u_{A1})\right]=\delta_2(u_{A1})
\left[{u_2(u_{A1})\over u_{\rm min}(u_{A1})}\right]^2\ ,
\end{equation}
where $u_{\rm min}(u_{A1})=u_{\rm min}[\eta_{\rm
min}(u_{A1}),u_{A1}]$. Figures \ref{eta_min} and
\ref{eta_min_delta2} show $\eta_{\rm min}$ as a function of
$u_{A1}$ and $\delta_2$, respectively. The critical value of $\eta$ is close to
1 for small values of $u_{A1}$ (or $\delta_2$) since, in this case,
the initial magnetic field is small, and almost all of the
internal energy needs to be lost in order for $B_{\rm max}/B_2$
($=u_2/u_{\rm min}$) to be large enough for $B_{\rm max}$
to reach equipartition.

As we noted in \S 1, there are direct observational indications
that at least in some relativistic shock sources the magnetic
field is not far below equipartition. Using the
calculated radiative shock structure, we can address the
question of how far from equipartition the magnetic field is at
the location where the bulk of the observed radiation is
emitted. A fluid element starts radiating just behind the shock
transition, where $B=B_2$ and $\delta=\delta_2$,
and as it flows downstream the magnetic field increases, and so does
the magnetic-to-thermal pressure ratio $\delta$.  Thus,
even if initially the magnetic field were much below equipartition
($\delta_2\ll 1$), it would be amplified by a large
factor ($\delta\gg\delta_2$) in the radiative zone and might
eventually even exceed equipartition ($\delta>1$). It is therefore conceivable
that a considerable fraction of the radiation could be emitted
from a region where the magnetic field is much
larger than the immediate-postshock value. We now examine this
possibility in a quantitative manner.

Since the internal energy of the electrons and positrons is
characterized by their thermal Lorentz factor $\gamma_e\propto
p_e/n_e\propto up_e$, their fractional remaining energy is given by 
$up_e(u)/u_2p_{e2}=u\tilde{p}_e(u)/u_2$.
Figure \ref{pu_delta2_combined}
shows this quantity as a function of $\delta_2$ at two different
instants: (i) at equipartition, when $\delta=1$, and (ii) when
$\delta=0.01$. For $\eta$ close to 1 a significant fraction of the
internal energy of the electrons and positrons is still left when the
field reaches equipartition even if $B_2\ll B_{\rm eq2}$. In this
case a significant portion of the radiation originates in a region
where the magnetic field is close to equipartition. We find that, so
long as the energy dissipated in the shock transition is deposited
mainly in the $e^\pm$ component (i.e., $\eta\approx 1$), then, even if
$\delta_2$ is as small as $\sim 10^{-4}-10^{-3}$, a considerable
fraction of the radiation is still emitted at $\delta>10^{-2}$. 
Note that, for $\eta\approx 1$ and $\delta_2\ll 1$, $a\sim 1/\delta_2\gg 1$
(see eq. [\ref{y2}]), so that, in the absence of a strong
external photon field ($b\ll a^{1/2}\sim\delta_2$), $y\approx
a^{1/2}\sim\delta_2^{-1/2}\gg 1$ (see eq. [\ref{y3}]). This
implies that SSC dominates the synchrotron emission, with
the synchrotron component contributing only a fraction 
$1/(1+y)\sim\delta_2^{1/2}\ll 1$ of the total radiation.
However, as we noted at the end of \S \ref{radiation} (see also
\S \ref{B1}), the synchrotron emission may dominate the SSC
contribution in the observed frequency range, in which case a
significant fraction of the observed  radiation could, in fact,
originate from a region where $\delta\gg\delta_2$.

In Figure \ref{syn_B} we show the fraction of
the total synchrotron emission that is emitted above a given
value of $B/B_{\rm eq}$, as well as the fraction of the
total emission contributed by the synchrotron process above that
value of $B/B_{\rm eq}$, as a function of the magnetic field
strength behind the shock. Three sets of parameters are
shown, labeled A through C, which correspond to the three limits
(from top to bottom, respectively) of equation (\ref{y3}). In case A,
SSC is dominant near the shock front, whereas further downstream the
magnetic field is strongly amplified by compression, and
synchrotron emission becomes dominant. Altogether, synchrotron
emission constitutes 45\% of the total emission, whereas
near the maximal field amplitude ($B\approx B_{\rm max}$) the 
synchrotron process contributes 94\% of the total emission. A large
fraction of the synchrotron radiation comes from a region where the
magnetic field is close to equipartition even though the magnetic
field just behind the shock transition is a factor of 10 below the
equipartition value ($\delta_2 \approx 10^{-2}$). In case B,
synchrotron emission is dominant and contributes 84\% of the
integrated total emission. The compression behind the shock is small
and the magnetic field is hardly amplified, so even near $B_{\rm max}$
the synchrotron fraction of the total emission (86\%) is
only slightly higher than the overall fraction.  In case C, the values
of $\eta$ and $u_{A1}$ are the same as in case A, and therefore
the initial and final values of the magnetic field are the
same. However, in case C the emission is dominated by IC
scattering of external photons ($b=100$), which are absent in case
A. Synchrotron emission now constitutes only 13\% of the
integrated total emission, but near $B_{\rm max}$ it contributes
59\% of the total. 
Most of the synchrotron emission comes from a
region where the magnetic field is close to equipartition --- a
larger fraction than in case A.  This is a reflection of the
fact that the ratio of synchrotron to total emission near
$B_{\rm max}$ relative to that near $B_2$ is larger in case C
because of the added cooling by IC scattering of the external photons.

\section{Applications}
\label{B1}

To assess the relevance of cooling-induced magnetic field
amplification to relativistic shock sources, one needs to
address the following questions: What are the likely values of
the normalized energy densities of the electrons/positrons and
the magnetic field ($\epsilon_e$ and $\epsilon_B$, respectively;
see \S 1) in the emission region? What are the expected values of
$\epsilon_e$ and $\epsilon_B$ immediately behind the shock
transition (which can be written, in terms of the shock
parameters defined in \S 2, as $\epsilon_{e2} = \eta$ and
$\epsilon_{B2} = \delta_2/3$)? Is the shock radiative? If the
shock is radiative but $\epsilon_{B2}$ is lower than the
observationally estimated value of $\epsilon_{B}$, can the
compressional amplification of the magnetic field in the
radiative zone make up the difference?

A strong relativistic shock ($u_1 \gg 1$, $u_{A1} \ll 1$) with a
transverse magnetic field satisfies $e_2 \approx 2u_1^2 w_1$ and
$B_2^2 \approx  8 u_1^2 B_1^2$ (see \S 2), implying that the
ratio of the magnetic energy density to the total matter energy
density is approximately constant across the shock and hence that
$\epsilon_{B2}$ can be expressed in terms of the enthalpy
density $w_1$ ($=\rho_1c^2$ for a ``cold'' upstream medium) and
the magnetic field amplitude $B_1$ ahead of the shock:
\begin{equation}\label{epsilon_B}
\epsilon_{B2} \approx {B_1^2\over 2\pi\rho_1 c^2}\ .
\end{equation}

As a concrete example of how the above considerations can be
applied, we again focus on GRBs and their afterglows. However,
at the end of this section we comment briefly on AGNs.
Starting with the $\gamma$-ray bursts themselves, we have
already observed (see \S 1) that efficiency considerations imply
that $\epsilon_e$ (in particular) and $\epsilon_B$ cannot be
much smaller than 1 in the emission region. The value of
$\epsilon_B$ may be constrained by observations through the
requirement that the cooling time does not exceed the
variability time scale $\delta t$. This translates into the
condition that the Lorentz factor of an electron that cools on
the time $\delta t$, $6\pi m_e c/\sigma_T
B^2\gamma\delta t (1+Y_2)$, must remain smaller than the
characteristic synchrotron Lorentz factor,
$(2\pi m_e c \nu_{\rm ob}/\gamma e B)^{1/2}$, that
corresponds to the observed frequency $\nu_{\rm ob}$.\footnote{To
simplify the discussion, we omit cosmological correction factors
from the expressions presented in this paper.} Following the
same steps as in the derivation of equation (\ref{check_IS}) and
again setting $\gamma_{12} = 2$, we deduce
\begin{equation}\label{epsilon_B_1}
\epsilon_B>8.3\times 10^{-8}(1+Y_2)^{-4/3}E_{52}^{-1}\gamma_{2}^{16/3}
t_{\rm dur,1}^{5/3}\zeta_{-2}^{2/3}
\left({h\nu_{\rm ob}\over 100\, {\rm keV}}\right)^{-2/3}
\end{equation}
(cf. \cite{SP97a}).  When the energy density of external photons is
negligible, one has $Y_2\approx y$ (see eq. [\ref{Y}] with $b=0$), so
that, if $y\gg 1$ (in which case $y \sim
(\epsilon_{e2}/\epsilon_{B2})^{1/2}$; see eqs. [\ref{y2}] and
[\ref{y3}]), the constraint (\ref{epsilon_B_1}) on $\epsilon_B$ would
be relaxed. However, $y\gg 1$ means that IC is dominant, with the
observed synchrotron radiation being only a fraction $\sim 1/y$ of the
total radiated energy. The requirement of a reasonable radiative
efficiency would then constrain $\epsilon_B$ to be not much smaller
than $\epsilon_e$ (e.g., Piran 1999).
The latter restriction would not arise if the observed
$\gamma$-ray emission represented the SSC rather than the
synchrotron component, but the synchrotron model appears to
account quite well for the observed spectra (e.g., \cite{L&P}),
whereas an SSC interpretation could entail a prohibitively large
energy input at the source.\footnote{For example, in the
model of Panaitescu and M\'esz\'aros (2000), the observed $\gamma$-ray
spectrum is identified with a singly scattered SSC
component. The characteristic thermal Lorentz factor of the
emitting electrons in this picture is $\sim 30$, which implies
that the peak synchrotron frequency is not very far from the
optical regime. There should thus be an optical pulse that
coincides with the prompt $\gamma$-ray emission, and the fact
that such a component has not been detected (e.g.,
\cite{Akerlof00}; \cite{Williams00}) indicates that the
Compton $y$ parameter is very large (Panaitescu and
M\'esz\'aros 2000 estimate $y \gtrsim 5000$).
Now, for the postulated thermal Lorentz factor, a typical seed photon
undergoes two scatterings before the Klein-Nishina
limit is reached. Since the emissivity ratios of the
synchrotron, singly scattered SSC, and doubly scattered SSC
components are $1:y:y^2$, it follows that the observed
$\gamma$-ray pulse may constitute only a fraction
$\sim 1/y\ll 1$ of the total burst emission (and hence that
the energy requirement on the source is a factor of $y$ larger).}

\begin{figure}
\begin{center}
  \plotone{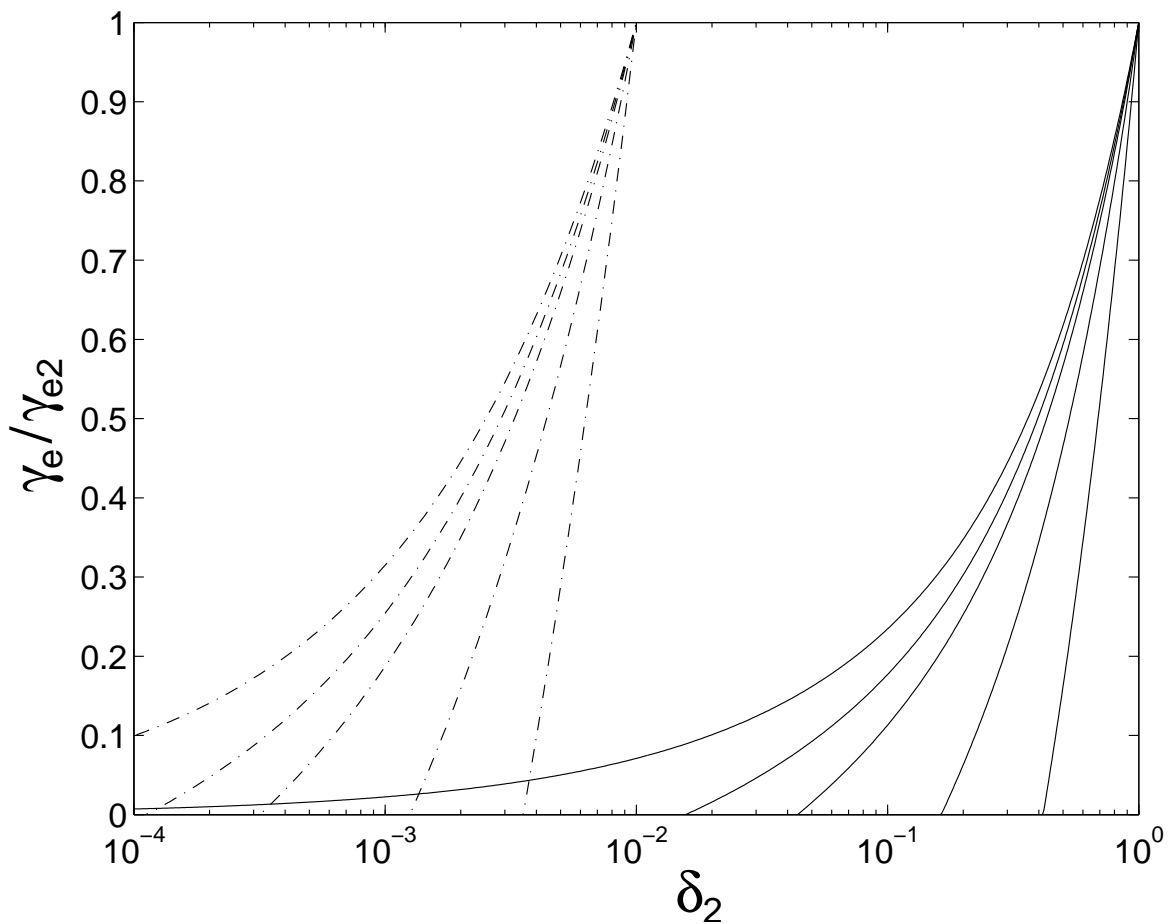} \figcaption[]{The fraction
    of the postshock internal energy still left in the
    electron/positron component as a function of the postshock
    magnetic-to-thermal pressure ratio, $\delta_2$: (i) at the point
    where the magnetic field attains equipartition with the thermal
    pressure, $\delta=1$ ({\it solid}\/ curves), and (ii) at the point where
    the magnetic pressure is $1\%$ of the thermal
    pressure, $\delta=0.01$ ({\it dash-dotted}\/ curves).  The curves in
    each of the two cases correspond to $\eta=0.5,0.75,0.9,0.95$, and
    1, respectively, from right to left.
\label{pu_delta2_combined}}
\end{center}
\end{figure}

\begin{figure}
\begin{center}
  \plotone{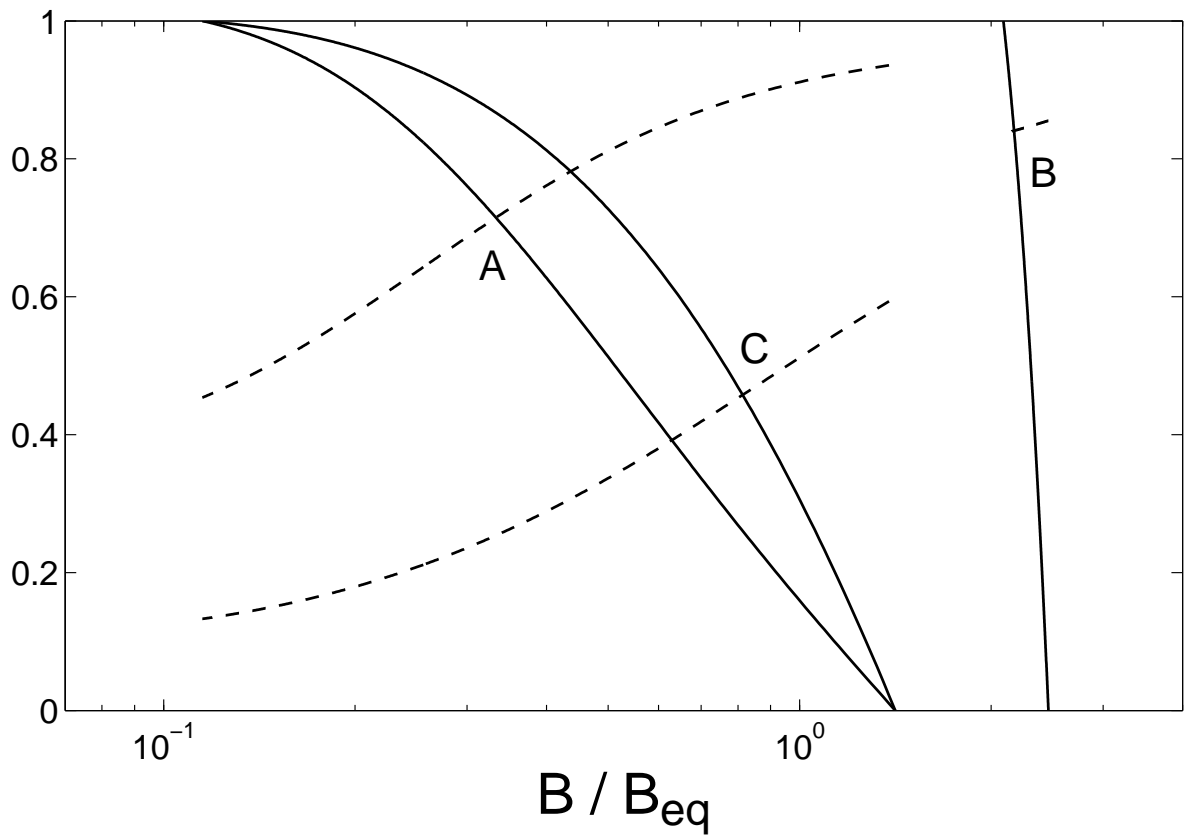} \figcaption[]{The {\it solid}\/ lines show the
    fraction of the synchrotron emission that is emitted in a
    region where the magnetic field is larger than a given value,
    $P_{\rm syn}(>B)/P_{\rm syn}({\rm total})$, as a function of
    the magnetic field strength (normalized by its equipartition
    value). The {\it dashed}\/ lines show the
    fraction of the total emission that is contributed by the synchrotron
    process, $P_{\rm syn}(>B)/P_{\rm total}(>B)$. The results are shown for
    three sets of parameters: (A)  $b=0$, $\eta=1$, $u_{A1}=0.03$; (B)
    $b=0$, $\eta=1$, $u_{A1}=1$; (C) $b=100$, $\eta=1$, $u_{A1}=0.03$.
\label{syn_B}}
\end{center}
\end{figure}

As we have already noted, the prompt $\gamma$-ray emission in GRBs is
believed to be the result of internal shocks within the flow. The
relevant preshock field is then the relic magnetic field advected from
the origin, and, to the extent that flux freezing is maintained,
$\epsilon_{B2} \propto B_1^2/\rho_1$ keeps the value that was
imprinted on the ejecta at the source.\footnote{Note in this
  connection that equation (\ref{epsilon_B}) shows quite generally
  that $\epsilon_{B2}$ will remain constant with time if the medium
  into which the shock propagates is either a supersonic outflow with
  a predominantly transverse magnetic field that is frozen into the
  matter (in which case both $B_1$ and $\rho_1^{1/2}$ scale inversely
  with distance from the center) or else is uniform (in which case the
  values of $B_1$ and $\rho_1$ are fixed). The former case applies to
  internal shocks and to the reverse shock (which propagate in the
  ejecta) as well as to a forward shock that propagates in a
  magnetized stellar-wind environment. The latter case corresponds to
  a forward shock in a uniform ISM. This shows that the predictions of
  conventional emission models of GRBs and their afterglows, which
  simply postulate a constant postshock value of $\epsilon_B$, will by
  and large continue to apply when this parameter is calculated
  self-consistently.} Spruit et al. (2001) pointed out that, under
these circumstances, values of $\epsilon_{B2}$ as high as $\sim 0.1-1$
may be expected in the commonly invoked GRB models that envision
large-scale, ordered magnetic fields actively driving the relativistic
outflow by tapping the rotational energy of the central source. In
particular, $\epsilon_{B2}$ could well be $\sim 1$ in collimated
(jet-like) outflows of this type. Spruit et al. further estimated
that, if the field is ``passive'' (in that it does not drive the flow
but is merely advected outward from the source), then $\epsilon_{B2}$
would lie in the range $\sim 10^{-6}-10^{-3}$. In the latter case,
compressional magnetic field amplification in the radiative zone could
increase the fraction of radiation going into the synchrotron
component (see Figs. \ref{pu_delta2_combined} and \ref{syn_B}).
For this mechanism to operate, it is necessary that the radiative
cooling time be much shorter than the dynamical time and that
$\epsilon_e$ be $\sim 1$. The first requirement should in general be
satisfied in internal shocks that produce GRBs (e.g., Piran 1999). The
conditions under which $\epsilon_e$ could be close to 1 are somewhat
uncertain, but there are several conceivable mechanisms for depositing
a significant fraction of the postshock internal energy in the
electron component even if the outflow consists predominantly of an
electron-proton plasma (e.g., \cite{B&M}). It should be even easier to
attain high values of $\epsilon_e$ if the plasma is dominated by
$e^\pm$ pairs, a situation that in fact arises naturally in some GRB
scenarios that involve magnetically driven outflows (e.g., Usov 1994;
\cite{MR}).

The long-lived afterglow emission of GRBs is most likely
associated with the forward shock that propagates into the
ambient medium. We have already noted in \S 1 that comparatively
high values of $\epsilon_e$ ($\sim 0.1-1$) and $\epsilon_B$
($\sim 0.01-0.1$) have been directly inferred from observations
of the GRB 970508 afterglow. 
These parameters have so far been less well constrained in other
sources, but in a recent modeling of four other afterglows
(\cite{PK01}) values in the range $10^{-2}\lesssim\epsilon_e\lesssim
10^{-1}$ and $10^{-6}\lesssim\epsilon_B\lesssim 10^{-1}$ have
been inferred. It is also worth noting in this connection the
suggestion by Galama et al. (1999) that comparatively low values of
$\epsilon_B$ may characterize afterglow sources that exhibit weak
radio emission (attributed by them to low values of the characteristic
frequency and peak flux of the synchrotron spectrum, both of which
scale as $\epsilon_B^{1/2}$).  The expected values of $\epsilon_{e2}$
and $\epsilon_{B2}$ in GRB afterglows depend on the GRB environment
(which directly affects such factors as the pair content, the
electron/positron acceleration efficiency, and the degree of
magnetization in the shocked gas).
The uniform-density model fittings of the GRB
970508 afterglow (\cite{W&G99}; \cite{GPS99}) yielded upstream
densities in the range $\sim 0.03-3\ {\rm cm}^{-3}$, which are typical
of a diffuse ISM. In the case of GRB 980329, a preshock density $\sim
10^3\ { \rm cm}^{-3}$, which is typical of a molecular cloud, has been
deduced (\cite{LCR}). In the Galactic ISM, the inferred magnitude of
the Alfv\'en speed does not vary strongly from diffuse to dense
environments and corresponds to values of $\epsilon_{B2}$ in the range
$\sim 3 \times 10^{-11}- 3 \times 10^{-10}$. If these values are also
typical of the host galaxies of GRBs, then the synchrotron emission
from their associated afterglows would in most cases be too weak to be
detected. However, all the afterglow sources observed so far
correspond to ``long'' bursts ($t_{\rm dur} \gtrsim 2\ {\rm s}$), 
for which a plausible origin is the collapse of a massive (e.g.,
Wolf-Rayet) star (e.g., Woosley 2000; M\'esz\'aros 2000). In this case,
the medium into which the forward shock expands could correspond to
the preburst wind from the central star (e.g., \cite{C&L00};
\cite{RR01}). It has been suggested (e.g., \cite{BC}) that winds from
Wolf-Rayet stars are magnetically driven and have spatially constant
Alfv\'en speeds corresponding to values of $\epsilon_{B2}$ that can be
as high as $\sim 10^{-4}$ . If such winds constituted the
environment into which the forward shock expands, then they could
account for afterglows with the lowest values of $\epsilon_B$ inferred to date
even if there were no additional magnetic field amplification
beyond that induced by the passage of the ambient field through
the shock transition.
Radiative cooling effects could increase the magnitude of
$\epsilon_B$ in the afterglow emission region above its value
immediately behind the forward shock. The cooling time behind
the forward shock is typically shorter than the dynamical time
during the early phase of the afterglow, but subsequently this
inequality is reversed. For a shock propagating into a
uniform-density ISM, the time of transition from radiative to
adiabatic evolution was calculated by Sari et al. (1998) and is given by
\begin{equation}\label{t_ad_ISM}
t_{\rm rad \rightarrow
ad}=4.6\, \epsilon_{B2}^{7/5}\epsilon_{e2}^{7/5}(1+Y_2)^{7/5}E_{52}^{4/5}
\gamma_{0,2}^{-4/5}n_1^{3/5}\ {\rm days}\quad {\rm (ISM)}\, ,
\end{equation}
where $\gamma_0$ is the initial Lorentz factor of the ejecta. In
the absence of a strong coupling between protons and
electrons behind the shock transition, $\epsilon_{e2}\approx 1$ is
required for the shock to be radiative. Equation
(\ref{t_ad_ISM}) indicates that, even if this condition is satisfied, 
$(1+Y_2)\epsilon_{B2}$ cannot be much smaller than 1 if the
radiative phase is to last through a significant fraction of the
early evolution of a typical afterglow (i.e., a few days or so). Taking
$\epsilon_{B2} \approx 10^{-10}$ as a representative ISM value,
we infer $(1+Y_2)\epsilon_{B2} \sim
(\epsilon_{e2}\epsilon_{B2})^{1/2}\sim 10^{-5}$, implying that
$t_{\rm rad \rightarrow ad}$ is only a few seconds! In fact, in this case,
even if the radiative phase lasted longer,
the expected magnetic field amplification in the
radiative zone would not be sufficient to raise $\epsilon_B$ to
a detectable level (see \S \ref{results}).

In the case of a stellar-wind environment, the ambient density
scales with distance $r$ from the source as $\rho(r)=A
r^{-2}$, where the constant $A\equiv\dot{M}/4\pi v_w$ is defined
in terms of the wind mass-loss rate $\dot{M}$ and speed $v_w$.
We are interested in the evolutionary phase during which the forward
shock has already decelerated significantly but is still
relativistic. Assuming that the mass is concentrated in a thin,
spherical shell of radius $R$ and Lorentz factor $\gamma$, we
can write $\gamma(R)=M/m(R)$, where $M$ ($=E/\gamma_0c^2$) is
the initial mass of the ejecta and $m(R)$ is the rest mass of the
swept-up matter (\cite{BM}; \cite{K&P97}). In our case $m(R)=4\pi
AR$ and therefore $\gamma(R)=L/R$, where $L=M/ 4\pi A$
is the distance from the center where the flow becomes nonrelativistic.
Using $t=R/4\gamma^2 c$, we obtain
\begin{eqnarray}\label{L}
\gamma=\left({E\over 16\pi A\gamma_0 c^3 t}\right)^{1/3}\, ,
\\
R=\left({E^2 t\over 4\pi ^2 A^2 \gamma_0^2 c^3}\right)^{1/3}\, .
\end{eqnarray}
The Lorentz factor of an electron that cools on a time scale $t$ is
given by
\begin{equation}\label{gamma_c}
\gamma_c={6\pi m_e c \over \sigma_T(1+Y_2)\gamma B^2 t}\ , 
\end{equation}
whereas the typical Lorentz factor of the electrons can be approximated by 
\begin{equation}\label{gamma_m}
\gamma_m\approx 610\epsilon_{e2}\gamma
\end{equation}
(e.g., Sari et al. 1998). Assuming that the preshock field $B_1= B_2/\gamma_1$
scales as $1/R$ (e.g., \cite{BC}), we find
$\gamma_c(t)\propto t^{2/3}$, so the ratio of the
radiative cooling time to the dynamical time ($t_{\rm
rad}/t_{\rm dyn}=\gamma_c/\gamma_m$) is $\propto t$. This shows
explicitly that at early times $t_{\rm rad}<t_{\rm dyn}$ and
the evolution is radiative. The transition
from the radiative to the adiabatic phase occurs approximately when
$\gamma_c/\gamma_m$ increases to 1.  Using $B^2/8\pi=\epsilon_B e$ and
$e=4\gamma^2\rho c^2$, we obtain
\begin{equation}\label{t_ad_wind}
t_{\rm rad \rightarrow
ad}=28.8\epsilon_{B2}\epsilon_{e2}(1+Y_2)A_*\  {\rm days}
{\rm \quad (Wind)}\ ,
\end{equation}
where $A_*\equiv A/(5\times 10^{11}\ {\rm g\ cm^{-1}})$ corresponds to
a stellar wind characterized by a speed of $10^8 \ {\rm cm\ s^{-1}}$
and a mass loss rate of $10^{-5}\ M_{\odot}\ {\rm yr}^{-1}$
(\cite{C&L00}). As can be seen from a comparison of equations
(\ref{t_ad_ISM}) and (\ref{t_ad_wind}), the radiative phase of the
forward shock will typically last longer in a wind environment than in
a uniform-density medium, which can be attributed (\cite{C&L00}) to
the higher ambient density encountered at early times by a shock that
propagates into a wind. Using equations (\ref{y2}) and (\ref{y3}) to
approximate $\epsilon_{e2} \epsilon_{B2} (1+Y_2) \lesssim
(\epsilon_{B2}/2)^{1/2}$ and setting $\epsilon_{B2} \approx 10^{-4}$
(the expected value in a strongly magnetized Wolf-Rayet outflow), we
infer from equation (\ref{t_ad_wind}) that $t_{\rm rad \rightarrow ad}
\lesssim 5A_*\ {\rm hr}$. During the radiative phase, a typical
electron could radiate up to $\sim 10\%$ of its postshock random
energy in a region with $\epsilon_B \gtrsim 10^{-2}$ (see Fig.
\ref{pu_delta2_combined}). However, the estimated value of $t_{\rm rad
  \rightarrow ad}$ is too small for the radiative field-amplification
mechanism to apply to the observations of a source like GRB 970508, in
which values of $\epsilon_B\gtrsim 10^{-2}$ have been inferred over
substantially longer time scales. Furthermore, even if the model were
applicable, this mechanism could not account for a value of
$\epsilon_B$ as high as 0.1 (the best-fit value for a stellar-wind
model; \cite{C&L00}) if $\epsilon_{B2}$ is not larger than $\sim
10^{-4}$.

Another potential source of detectable emission in GRB sources is the
reverse shock, which runs into the ejecta shell when the latter starts
to be decelerated by the inertia of the swept-up ambient gas. In fact,
a shock of this type is believed to be the source of the prompt
optical emission (or ``optical flash'') that was observed in GRB
990123 (\cite{Akerlof99}; Sari \& Piran 1999a,b; \cite{MR99}) as well
as of a rapidly decaying ``radio flare'' detected about a day
after the burst in this source (and evidently also present in
GRB 990123; Kulkarni et al. 1999). So far, however, these
observations have not provided strong constraints on the parameters
$\epsilon_e$ and $\epsilon_B$ in the emission region.
It may, however, be expected that, since the reverse shock propagates
into the ejecta, it is characterized by values of $\epsilon_{e2}$ and
$\epsilon_{B2}$ that are similar to those in internal shocks.

GRB 990123 was a comparatively long $\gamma$-ray burst: such sources
likely possess a relatively thick ejecta shell (of thickness $\Delta$
as measured by a stationary local observer), and their prompt optical
emission should peak at $t_{\rm peak} \approx \Delta/c$, the shock
crossing time of the shell (which is approximately equal to the
duration of the burst), by which time the reverse shock would be
relativistic (e.g., \cite{SP99b}). This is a natural time scale to use
in a comparison with the cooling time for determining whether the
shock is radiative. We can repeat the procedure applied above to the
forward shock by comparing the characteristic Lorentz factors
$\gamma_m$ and $\gamma_c$, except that in this case we substitute for
$\gamma$ in equation (\ref{gamma_c}) for $\gamma_c$, by the Lorentz
factor $\gamma_3$ of the shocked shell material, which is given in
terms of the Lorentz factor $\gamma_{\rm sh}$ of the un-shocked
portion of the ejecta shell (both measured in the stationary frame) by
\begin{equation}\label{gamma_3}
\gamma_3 = \gamma_{\rm sh}^{1/2}f^{1/4}/\sqrt{2}
\end{equation}
(e.g., \cite{SP95}), where
\begin{equation}\label{f}
f = {E \over 4\pi\gamma_{\rm sh}^2 R_\Delta^2 \Delta n_{p1} m_p c^2}
\end{equation}
is the ratio of the proper mass densities in the shell and the
ambient medium (assumed to be dominated by protons) and
$R_\Delta = \gamma_3^2\Delta$ is the radius where the reverse
shock crosses the shell. Similarly,
we replace $\gamma$ in equation (\ref{gamma_m}) for $\gamma_m$
by the relative Lorentz factor between the shocked and the
unshocked shell material, $\bar \gamma_3 = \gamma_3/\sqrt{f}$.

In the case of a uniform ambient medium, the condition
$\gamma_c(t_{\rm peak})/\gamma_m(t_{\rm peak})\le 1$ implies
\begin{equation}\label{epsilon_B_epsilon_e_ISM}
\epsilon_{e2} \epsilon_{B2} (1+Y_2) \ge 0.12\,
E_{52}^{-1/4}n_{p1}^{-3/4}(t_{\rm peak}/50\, {\rm s})^{-1/4}
(\gamma_{\rm sh}/200)^{-1}\, ,
\end{equation}
where we have normalized $t_{\rm peak}$ and $\gamma_{\rm sh}$ by
their estimated values in GRB 990123
(\cite{SP99a}).\footnote{The condition $\gamma_c=\gamma_m$ is
probably a bit conservative. For one thing, as we have demonstrated
in \S 3, the cooling induces compression in the radiative layer,
which has the effect of enhancing the radiative
losses. Furthermore, the radiative efficiency decreases only
gradually after $\gamma_c$ comes to exceed $\gamma_m$: in fact,
the fraction of the internal electron/positron energy that is
radiated by the synchrotron process after that time is given by
$(\gamma_m/\gamma_c)^{p-2}$, where $p$ (which has a canonical
value of 2.5) is the power-law index of the radiating particle distribution
(e.g., \cite{MSB}). Thus, the right-hand
sides of the inequalities
(\ref{epsilon_B_epsilon_e_ISM})--(\ref{epsilon_B2_WIND})
are probably only upper limits.}
Assuming $\epsilon_{e2} \sim 1 \gg \epsilon_{B2}$ [which implies, in the
absence of an external radiation field,
$\epsilon_{e2} \epsilon_{B2} (1+Y_2) \lesssim
(\epsilon_{B2}/2)^{1/2}$], we obtain a lower bound on
$\epsilon_{B2}$,
\begin{equation}\label{epsilon_B2}
\epsilon_{B2} \gtrsim 0.03\,
E_{52}^{-1/2}n_{p1}^{-3/2}(t_{\rm peak}/50\, {\rm s})^{-1/2}
(\gamma_{\rm sh}/200)^{-2}\, .
\end{equation}
The inequality (\ref{epsilon_B2}) could in principle be satisfied if the GRB
outflow is driven magnetically, in which case the ejecta could
be characterized by $\epsilon_{B1} \sim \epsilon_{B2} \sim 1$.
However, even if
$\epsilon_{B2}$ is near its estimated lower bound, compressional
amplification could boost the magnetic field to
equipartition values in the radiative zone (see
Figs. \ref{eta_min_delta2} and \ref{pu_delta2_combined}).
The corresponding condition for a stellar wind environment
(assuming strong hydrogen depletion as appropriate for a
Wolf-Rayet progenitor) is
\begin{equation}\label{epsilon_B_epsilon_e_WIND}
\epsilon_{e2} \epsilon_{B2} (1+Y_2) \ge 9.6 \times 10^{-6}
E_{52}^{1/2}A_*^{-3/2}(t_{\rm peak}/50\, {\rm s})^{1/2}
(\gamma_{\rm sh}/200)^{-1}
\end{equation}
(\cite{C&L00}).  For $\epsilon_{e2} \sim 1 \gg
\epsilon_{B2}$ and a negligible external radiation field,
equation (\ref{epsilon_B_epsilon_e_WIND}) implies
\begin{equation}\label{epsilon_B2_WIND}
\epsilon_{B2} \gtrsim 1.8 \times 10^{-10}
E_{52}A_*^{-3}(t_{\rm peak}/50\, {\rm s})
(\gamma_{\rm sh}/200)^{-2}\, ,
\end{equation}
which demonstrates that the reverse shock is
always radiative in this case for plausible parameter values.
It is worth noting, however, that if the
reverse shock is radiative at the time that it crosses the
shell, then it will rapidly cool and become invisible at later
times (e.g., \cite{SP99a}). Since this was not the
case for the optical flash in GRB 990123, we can
conclude, on the basis of equations (\ref{epsilon_B_epsilon_e_WIND}) and
(\ref{epsilon_B_epsilon_e_ISM}), that the outflow in this
source does not expand into a wind (as was already inferred by
\cite{C&L00}) and that the value of $\epsilon_{e2}\epsilon_{B2}$
behind the reverse shock is $< 0.1$.

We can summarize our discussion of GRB sources as follows. The field
amplification mechanism discussed in this paper could in principle
operate in shocks that are driven into the ejecta: either the internal
shocks that give rise to the prompt $\gamma$-ray emission, or the
reverse shock that produces the 
optical flash and radio flare emission. Except for
reverse shocks in sources with a uniform ambient medium in the case
when the condition (\ref{epsilon_B_epsilon_e_ISM}) is not satisfied,
these shocks will be radiative if most of the energy dissipated in the
shock transition is deposited into the electron/positron component.
Compressional field amplification in the radiative zone of these
shocks could enable a significant fraction ($\sim 0.1-0.3$) of the
radiated energy to be emitted from regions where the field is
substantially stronger (by up to $\sim 1-2$ orders of magnitude in
$\epsilon_B$) than immediately behind the shock transition.  If the
preshock magnetization is comparatively weak ($\epsilon_{B1}\ll 1$),
the radiative amplification could significantly enhance the
synchrotron emission and may in certain cases help to bring it above
the detection limit.  If the ejecta is already moderately strongly
magnetized ($\epsilon_{B1}\gtrsim 10^{-2}$), this process could
increase the field in the emission region to equipartition values.

We have also concluded that the compressional amplification mechanism
will not significantly affect the afterglow emission of GRB sources,
both because the forward shock typically undergoes a transition to
adiabatic evolution early on, and because, even during the radiative
phase, the expected amplification falls short of the required
enhancement for standard ISM environments.  We noted that magnetized
stellar winds from Wolf-Rayet stars could provide an adequate
environment in which afterglows with values of
$\epsilon_B$ that lie near the low end of the observationally
inferred range would be produced in adiabatic
shocks without requiring further field amplification.
However, the comparatively high ($\gtrsim 10^{-2}$) values of
$\epsilon_{B}$ indicated in a source like GRB 970508 cannot be
accounted for by the basic shock model and typical ISM parameters. We
do not pursue this topic further in this paper, but we note that
several suggestions that bear on the need to have high values of
$\epsilon_B$ and $\epsilon_e$ in such shocks have already been made in
the literature (e.g., \cite{ML}; \cite{SU}; \cite{TM}; \cite{LB01};
\cite{KG01}).

The observational data on spectral flares in AGNs are not as extensive
as in GRB sources, and, correspondingly, their theoretical study has
so far been less developed. However, as we noted in \S 1, the blazar
flares have been similarly interpreted in terms of nonthermal emission
from relativistic, and likely highly collimated, outflows from the
vicinity of a central compact object.\footnote{One key difference
between the high-energy flares in AGNs and GRB sources is in their
observed duration, which is of the order of hours in blazars (e.g.,
\cite{W98}; \cite{M99}) as compared with seconds in GRBs. A related
difference is in the inferred bulk Lorentz factor of the outflow:
$\lesssim 10$ in blazars vs. $\gtrsim 10^2$ in GRBs.} In particular,
the variable emission has been modeled in terms of internal shocks in
a magnetized relativistic outflow, possibly formed from overtaking
collisions of disturbances (``shells'') in the flow (e.g.,
\cite{MG95}; \cite{RL97}; \cite{SMMP}; Levinson 1998, 1999). In these
models, the high-energy emission is attributed to an SSC or an ERC
component. The quasi-isotropic external radiation field invoked in the
ERC interpretations can arise from the nuclear accretion disk (e.g.,
\cite{D&S}), or, perhaps more likely, from scattering by the broad
emission-line region (BELR) ``clouds'' of the nuclear continuum
radiation (e.g., \cite{SBR}; \cite{BL95}) or of the beamed jet
emission itself (\cite{GM}), as well as from the near-infrared
emission of warm dust outside the BELR (\cite{BSMM}). As we discussed
in \S \ref{results}, if the emission from a radiative shock is
dominated by IC scattering of external photons, then the synchrotron
radiation component of the shock will have a relatively stronger
contribution from regions of high values of $B/B_{\rm eq}$ (see Fig.
\ref{syn_B}).\footnote{The emission characteristics of GRB sources
might also be affected by the interaction of the outflow with an
external radiation field. For example, Lazatti et al. (2000) considered the
possibility that the prompt $\gamma$-ray emission arises in the
course of the Compton-drag deceleration of a relativistic outflow as
it propagates through the debris of a progenitor massive star; in
this scenario, magnetic fields play no role in the emission process.
On larger scales, Madau, Blandford, \& Rees (2000) proposed that
backscattering of the GRB radiation by the comparatively dense
massive-star environment could lead to a delayed MeV emission,
again through the Compton-drag effect on the bulk flow. In the
latter case, ERC emission by a ``hot'' (i.e., relativistic)
electron/positron component of the flow could in principle also
contribute to the observed radiation (cf. the analogous proposal
for blazars by \cite{GM}). So far, however, there has been no
discussion in the literature of the possible appearance of an
ERC emission component in GRB sources.} Most of the blazar
emission models considered to date have, however, taken $\gamma_m$ to
lie below $\gamma_c$ (e.g., \cite{MGC92}; Sikora et al. 1994;
but see \cite{S01}) and thus
correspond to nonradiative shocks. It is, nonetheless, likely that
the cooling times could be quite short in these sources, as suggested,
for example, by very rapid flares in blazars like 3C 279 (e.g.,
\cite{W98}), Mrk 421 (e.g., \cite{M99}), and AO 0235+164 (e.g.,
\cite{K99}). Furthermore, there are indications that AGN jets may
contain a significant, or even dominant, electron/positron component
(e.g., \cite{RFCR}; \cite{WHOR}; \cite{SM00}), which would naturally
lead to high values of $\epsilon_e$ in the shocks. Given that the
outflows are likely magnetically driven (e.g., \cite{BZ77};
\cite{BP82}), comparatively high values of $\epsilon_B$ may also be
expected. It is thus conceivable that at least some of the synchrotron
emission from these sources originates in radiative shocks in which
the magnetic field is amplified by cooling-induced compression.

\section{Conclusions}
\label{conclusions}

Gamma-ray bursts and their afterglows, as well as energetic flares
from blazars and miniquasars, are commonly interpreted in terms of
shocks in a relativistic outflow from a compact object that emit
nonthermal (synchrotron and inverse Compton) radiation. In this paper
we have investigated the structure of radiative shocks in such
sources. A shock is ``radiative'' if the emission mechanism can tap
into the bulk of the postshock internal energy and if the radiative
cooling time behind the shock transition is shorter than the relevant
dynamical time (the characteristic time for adiabatic energy losses).
Such a shock is characterized by a radiative zone of finite width
where most of the energy dissipated in the shock transition front is
radiated away. The structure of radiative shocks has been previously
studied in the nonrelativistic regime in the context of the
interstellar medium, where the dominant emission processes are
thermal. One particularly interesting finding of these studies has
been the strong effect that such a shock could have on the normal
component of a magnetic field that is frozen into the gas, and, in
turn, the potentially important role that the field could play in
limiting the compression in the shock. When the shock is nonradiative,
the compression ($n_2/n_1$) and resultant magnetic field amplification
($B_2/B_1=n_2/n_1$) have a value of at most a few in the
nonrelativistic regime (generalizing to a few times the shock Lorentz
factor $\gamma_1$ in the relativistic case, with $n$ and $B$ measured
in the fluid rest frame).  Efficient cooling can lead to a strong
enhancement of the density and the field amplitude in the radiative
zone, but when the magnetic pressure attains equipartition with the
thermal pressure, further compression is inhibited and subsequent
cooling proceeds at nearly constant density. We have shown that, when
synchrotron radiation dominates the cooling, the interplay between the
cooling and the compression becomes more pronounced on account of the
feedback effect between the field amplification and the emission
process: a stronger magnetic field increases the emissivity, which in
turn induces a larger compression that further amplifies the field.

The cooling-induced compressional field amplification may have
potentially significant consequences in sources that harbor radiative
shocks: if the preshock magnetic field is low, this mechanism could in
some cases increase it to a level where the synchrotron radiation
becomes detectable, and if the preshock field is already moderately
strong (so that it is within an order of magnitude of equipartition in
the shocked but uncooled gas), this process could result in a
considerable fraction ($\gtrsim 0.1$) of the radiation being emitted
from regions where $B \approx B_{\rm eq}$. When the shock is radiative
but the field immediately behind the shock transition is below
equipartition, the initial cooling would be dominated by the
inverse-Compton process: either synchrotron self-Compton (SSC) or
external-radiation Compton (ERC), depending on which seed-photon
energy density is higher in the shock frame. In this case, a larger
fraction of the synchrotron emission comes from regions with a
stronger magnetic field if ERC dominates SSC.

For standard scenarios of GRB outflows and their environments, this
field amplification mechanism may be relevant to internal shocks that
produce the prompt $\gamma$-ray emission and possibly also to the
reverse shock responsible for the optical ``flash'' and
radio ``flare'' in sources like GRB 990123.
It will likely not be
of much relevance to the forward shock that gives rise to the GRB
afterglow, although we pointed out that if the shock expands into a
magnetized wind from a Wolf-Rayet star then no further field
amplification would be required to account for the lowest $B/B_{\rm
eq}$ values that have so far been inferred from afterglow
observations.  We also suggested that radiative shocks in which this
mechanism operates could occur in blazars.

\acknowledgments We are grateful to Tsvi Piran for many valuable
conversations. JG acknowledges a Priscilla and Steven Kersten
Fellowship at the University of Chicago and thanks the Department of
Astronomy and Astrophysics at the University of Chicago for its
hospitality. AK acknowledges a Forchheimer Fellowship and thanks the
Racah Institute at the Hebrew University for hospitality. This
research was supported in part by the US--Israel BSF grant
BSF-9800225 (JG) and by NASA grant NAG 5-9063 (AK).

\end{document}